\documentclass[10pt]{article}
\usepackage[utf8]{inputenc}
\usepackage[margin=0.9in]{geometry}
\usepackage[affil-it]{authblk}
\usepackage[normalem]{ulem}
\usepackage{bm}
\usepackage{amsmath}
\usepackage{float}
\usepackage{graphicx}
\usepackage{amssymb}
\usepackage[breaklinks=true,colorlinks=true,hypertexnames=false,linktoc=section,pdfencoding=auto,psdextra=true,linkcolor=blue,anchorcolor=black,citecolor=blue,filecolor=black,menucolor=black,runcolor=black,urlcolor=blue]{hyperref}
\usepackage{microtype}
\usepackage{cite}
\usepackage{subcaption}
\usepackage{placeins}
\usepackage{mathtools}
\usepackage{csquotes}
\usepackage{comment}
\usepackage{multirow}
\usepackage{array}
\usepackage{enumitem}
\usepackage[ruled,vlined]{algorithm2e}
\usepackage{algorithmic}

\usepackage[dvipsnames]{xcolor}

\providecommand{\keywords}[1]{\textbf{\textit{Keywords---}} #1}

\title{Multiscale nonlocal beam theory: An application of distributed-order fractional operators}
\author[1]{Wei Ding\thanks{To whom correspondence should be addressed. Email: ding242@purdue.edu or fsemperl@purdue.edu}}
\author[1]{Sansit Patnaik}
\author[1]{Fabio Semperlotti${}^*$}
\affil[1]{Ray W. Herrick Laboratories, School of Mechanical Engineering, Purdue University, West Lafayette, IN 47907}

\begin{document}
\date{}
\maketitle
\vspace*{-1cm}
\begin{abstract}
This study presents a comprehensive theoretical framework to simulate the response of multiscale nonlocal elastic beams. By employing distributed-order (DO) fractional operators with a fourth-order tensor as the strength-function, the framework can accurately capture anisotropic behavior of 2D heterogeneous beams with nonlocal effects localized across multiple scales. Building upon this general continuum theory and on the multiscale character of DO operators, a one-dimensional (1D) multiscale nonlocal Timoshenko model is also presented. This approach enables a significant model-order reduction without compromising the heterogeneous nonlocal description of the material, hence leading to an efficient and accurate multiscale nonlocal modeling approach. Both 1D and 2D approaches are applied to simulate the mechanical responses of nonlocal beams. The direct comparison of numerical simulations produced by either the DO or an integer-order fully-resolved model (used as ground truth) clearly illustrates the ability of the DO formulation to capture the effect of the microstructure on the macroscopic response. The assessment of the computational cost also indicates the superior efficiency of the proposed approach.
\vspace*{0.3cm}

\noindent\keywords{Multiscale modeling, Nonlocal elasticity, Distributed-order operators, Layered structures}
\end{abstract}

\begin{figure}[ht]
\renewcommand{\figurename}{Graphical Abstract}
\renewcommand{\thefigure}{}
    \centering
    \includegraphics[width=0.6\textwidth]{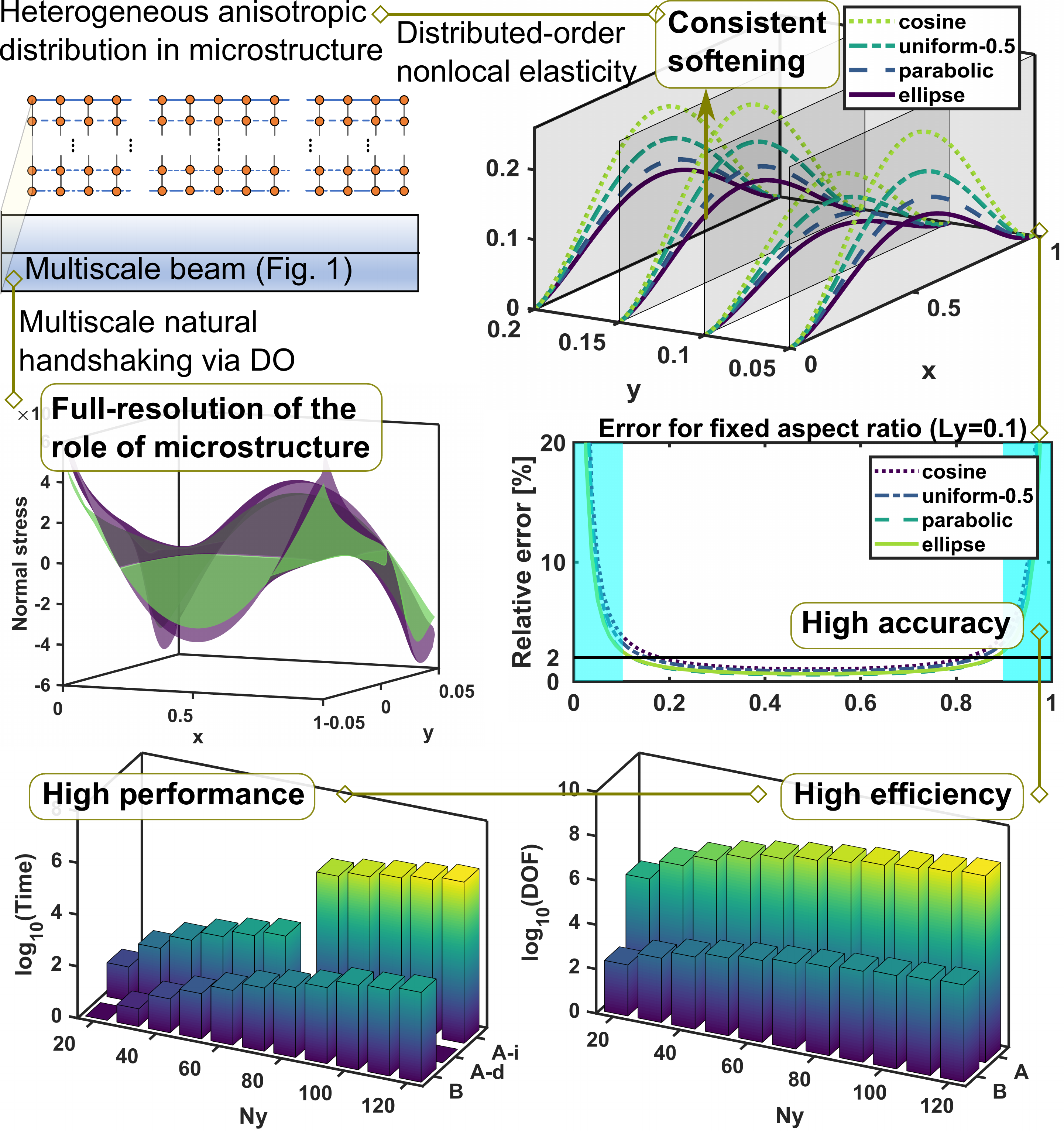}
    \caption{Impact of distributed-order nonlocal elasticity modeling on multiscale mechanics.}
\setcounter{figure}{1}
\end{figure}

\section{Introduction}
\label{sec: introduction}
The rapid development of novel manufacturing techniques has greatly accelerated the discovery and fabrication of complex materials including, but not limited to, composites \cite{fish2000multiscale,fu2022novel}, metamaterials \cite{celli2018pathway,kadic20193d,liu2020big}, and functionally graded materials~\cite{parameswaran2000processing,stempin2021formulation}. While applications can span a diverse range such as wave-guiding~\cite{liu2018tunable}, sensors and micro/nano-electromechanical devices \cite{george2010atomic}, and even biological implants \cite{wang2016topological}, all these materials are characterized by highly heterogeneous compositions and architectures. As complex material configurations and microstructural properties become progressively more accessible, the ability to exploit them in practical applications requires access to modeling techniques that can effectively capture the complex nature of these material systems and provide a viable route to accurate and efficient modeling.

Over the past decade, several theoretical and experimental investigations have highlighted the significance of multiscale and nonlocal effects for the accurate evaluation of the elastic response of the above-mentioned classes of materials \cite{van2020roadmap,shaat2020review}. The origin of nonlocal effects has been primarily attributed to the presence of structural heterogeneity and long-range interactions at the microstructural scales \cite{eringen1972nonlocal}. For the most part, their impact was studied and considered prominent for analyses conducted at a given scale. However, recent studies focused on the deformation of heterogeneous solids (e.g. porous solids \cite{patnaik2022role, sciarra2007second, dazel2007alternative, fellah2003measuring}, granular solids \cite{gonzalez2012nonlocal,misra2017elastic,giorgio2020chirality}, composites \cite{silling2014origin,trovalusci2014particulate,steigmann2015two}, functionally-graded solids \cite{stempin2021formulation,patnaik2021variable}, lattice structures \cite{gonella2008homogenization,russillo2022wave}, metamaterials \cite{willis2011effective,madeo2016first,mei2022nonlocality}), and even intentionally designed nonlocal structures \cite{zhu2020nonlocal,nair2019nonlocal,nair2022nonlocal} have demonstrated that nonlocal effects can also originate and localize at the meso- and macro scales. These studies reinforced the realization that nonlocal effects can exist and, more importantly, can interact across scales. This complex multi-level nonlocal phenomenon was the focus of a recent study \cite{ding2022multiscale} in which the phenomenon was denominated ``multiscale nonlocal elasticity". 

Real-world applications that exemplify the characteristics and the importance of multiscale nonlocal elasticity include, for example, layered and woven composites \cite{chakraborty2004spectrally,fu2022novel}, 3D printed solids \cite{monaldo2021multiscale,fu2022novel}, functionally graded solids \cite{parameswaran2000processing,stempin2021formulation}, solids with graded porosity \cite{sankar2016effect,mannan2017correlations,mannan2018stiffness}, and semiconductors fabricated by atomic layer deposition \cite{george2010atomic}. In order to support the discovery, the understanding, and the performance assessment of these material systems, it is essential to develop computationally efficient multiscale models that offer two fundamental characteristics \cite{hoekstra2014multiscale,chopard2014framework}: C1) the ability to simultaneously capture nonlocal effects across scales within a physically consistent and mathematically well-posed framework, and C2) the ability to naturally interconnect scales and exchange information without resorting to artificial hand-shaking approaches. From a practical perspective, these features enable capturing a range of multiscale nonlocal effects such as global softening or stiffening, localized softening or stiffening, displacement distortion, anomalous dispersion, energy concentration, and surface effects \cite{askes2011gradient,polizzotto2018anisotropy,russillo2022wave,patnaik2022role,ding2022multiscale}. 

Over the past 50 years, several efforts have been made to develop multiscale approaches to simulate the response of complex materials. From a very high-level perspective \cite{ding2022multiscale}, these multiscale approaches can be broadly classified as molecular dynamics models \cite{rapaport2004art}, local continuum models \cite{fish2000multiscale,willis2011effective,weinan2011principles}, and nonlocal continuum models \cite{eringen1972nonlocal,askes2011gradient,silling2000reformulation}. While the existing classes of multiscale approaches have been able to address several different aspects that characterize the response of multiscale solids, they are not equipped to capture multiscale nonlocality \cite{ding2022multiscale}. Of all the different classes of methods, molecular dynamics models are likely to enable the most accurate approach via a direct resolution of the different nonlocal material scales. However, the large number of degrees of freedom typical of molecular models renders their practical application for macro-scale analyses infeasible. On the other hand, local continuum approaches offer good computational efficiency at the macro scales but cannot capture nonlocal effects unless fully resolving the geometry of the microstructure (which clearly comes at the expense of computationally prohibitive resources \cite{weinan2011principles,patnaik2022role}). Finally, existing continuum nonlocal models can capture nonlocal effects but they are typically localized at a single material scale. In fact, the majority of existing nonlocal models directly embed the nonlocal interactions within the continuum level description via nonlocal material parameters (often obtained using a phenomenological approach) resulting in the so-called \textit{implicitly} multiscale models \cite{mcdowell2010perspective}. This implicit and artificial handshaking strategy encounters serious difficulties when enforcing thermodynamic balance principles in a strong sense across the multiple material scales \cite{polizzotto2001nonlocal,rivarola2017thermodynamic}.

Very recently, the authors showed in \cite{ding2022multiscale} that distributed-order fractional calculus (DO-FC) presents the most natural strategy to account for the multiscale nonlocal behavior within a generalized elasticity model. The distributed-order (DO) approach also allows addressing the previously mentioned limitations of existing multiscale approaches. DO-FC is a natural generalization of constant order fractional calculus (CO-FC) that offers unique features for the development of multiscale (nonlocal) modeling approaches. To-date, the main applications have been in the fields of viscoelasticity \cite{suzuki2021anomalous,failla2020advanced}, anomalous transport \cite{atanackovic2009time1,sandev2015distributed}, and control \cite{caputo1995mean,fernandez2017asymptotic}. A detailed review of DO-FC and its application to the analysis of real-world multiscale systems can be found in \cite{ding2021applications}. In the present study, we leverage the multiscale characteristic of DO operators to capture the response of multiscale beams by adopting the multiscale framework developed in \cite{ding2022multiscale}.


\subsection{The distributed-order approach to multiscale nonlocal material modeling}
\label{ssec: DO_theory_review}
In this section, in an effort to motivate the use of the DO-FC based multiscale nonlocal theory developed in \cite{ding2022multiscale} to model multiscale beams, we summarize the major highlights of the corresponding theory. More specifically, we briefly discuss how DO calculus and the overall theory in \cite{ding2022multiscale} naturally enable the previously discussed characteristics (C1 and C2) of multiscale models. The different physical mechanisms and concepts introduced in the discussion are schematically illustrated in Fig.~\ref{fig: Overview}(a). 

Firstly, in regards to achieving C1, note that the DO operators automatically inherit the nonlocal properties of CO-FC, since they are obtained following an integration of the power-law kernel of CO operators over an extended range of orders \cite{caputo1995mean}. More importantly, with the multiple co-existing orders being stacked together, the DO operator can be used naturally to capture heterogeneous nonlocal effects localized across multiple co-existing material scales (where the nonlocal effects corresponding to a specific material scale is represented by the corresponding CO strength \cite{patnaik2020generalized,alotta2022unified}). This latter concept was leveraged in \cite{ding2022multiscale} to develop a distribution of CO nonlocal stress-strain constitutive relations that captures the localized CO nonlocal effects at the different material scales. The positive-definite nature (and hence, the mathematical well-posedness) of the model was guaranteed by the symmetric power-law kernel \cite{polizzotto2001nonlocal} while its ability to capture structural anisotropy was enabled by the use of a fourth-order strength function tensor \cite{ding2022multiscale}. 

Finally, in regards to achieving C2, note that the strength function, when evaluated for a specific CO, captures the net contribution of the corresponding scale-specific nonlocal effects within the overall multiscale nonlocal phenomenon \cite{ding2021applications, ding2022multiscale}. More specifically, the strength function of the DO derivative, that (mathematically) serves as a order-weighting function for the range of CO (nonlocal effects), provides a coherent mathematical basis that is required for a physically consistent representation of the natural nonlocal interactions between the localized scale-specific nonlocal effects \cite{ding2022multiscale}. In equivalent terms, the strength-function serves as a tool to achieve an explicit, fully-resolved, and natural handshaking (or interaction) of the multiple (nonlocal) material scales (thanks also to the thermodynamic consistency of the multiscale model following C1). We will explore the latter concept in more detail in this study.

\subsection{Problem setup: Objectives and broader relevance of the study}
\label{ssec: Problem_setup}
In this study, we showcase the application of the DO nonlocal elasticity theory (reviewed above in \S\ref{ssec: DO_theory_review}) to perform accurate and efficient simulations of 2D heterogeneous beams with multiscale anisotropically-nonlocal effects. The study is constructed and presented through a benchmark structural analysis problem that highlights the unique features of DO-FC for multiscale nonlocal analysis. This benchmark problem consists in assessing the static response of a 2D beam characterized by a heterogeneous microstructure. In order to better communicate the objectives of the study, we first introduce the specific benchmark problem and present a high-level discussion on its relevance to the general problem of multiscale modeling.

\begin{figure}[ht!]
    \centering
    \includegraphics[width=1\textwidth]{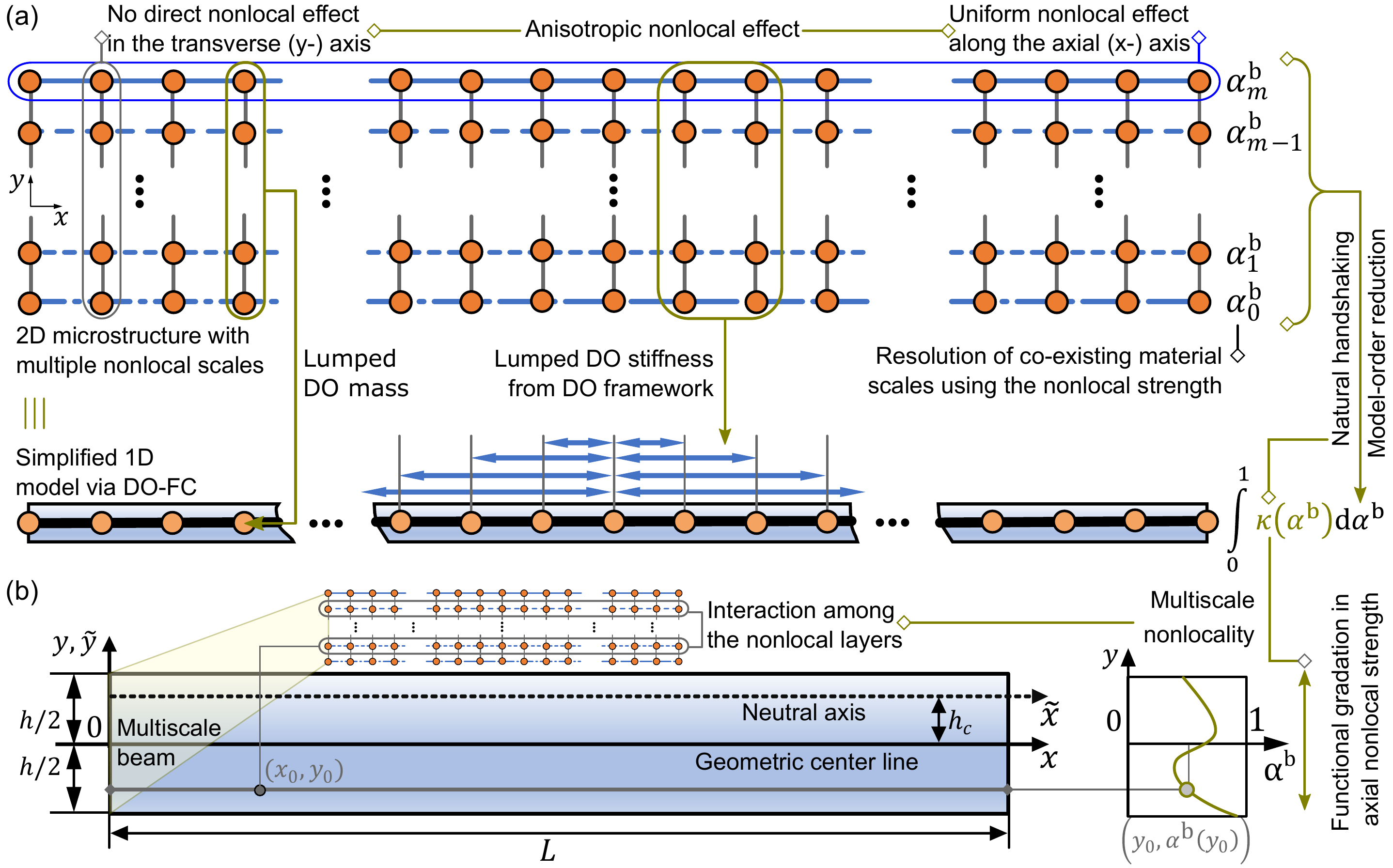}
    \caption{Schematics of a 2D heterogeneous beam with multiscale anisotropically-nonlocal effects. (a) Multiscale mechanics via DO calculus can capture the detailed 2D microstructure with multiple nonlocal scales. The microstructure consists of $m$ layers (in the $y$-direction) of discrete lattice chains that exhibit nonlocal effects along the $x$-axis. The nonlocal heterogeneity across different layers, labeled by $\alpha_i^\mathrm{b}$, $\forall i \in {0,1,...,m}$, is schematically illustrated by blue lines (in different line styles). By stacking the heterogeneous nonlocal order in the transverse ($y$) direction into a single DO operator, the 2D microstructure can be reduced naturally into a simplified 1D DO model while still preserving the coexistence of nonlocal effects across multiple scales. (b) Schematic of a benchmark problem showing the corresponding continuous DO representation of the beam with axial length $L$ and transverse thickness $h$. The geometry of the beam is defined within the $x$-$y$ coordinates ($x$ on axial direction and $y$ on transverse direction). For a given point $(x_0,y_0)$ within the elastic beam, the nonlocal interaction only arises from other points at the same transverse position $y_0$. It follows that nonlocality is homogeneously distributed along the axial direction but heterogeneously distributed along the transverse direction (which is consistent with the heterogeneous and anisotropic nonlocal configuration in (a)). The curve in the right $y$-$\alpha^{\textrm{b}}$ coordinates shows the heterogeneous distribution of order $\alpha^{\textrm{b}}(y) \in [0,1]$ that arises from the discrete order $\alpha_i^\mathrm{b}$ in the 2D microstructure. Physical points at the same transverse position $y_0$ experience the same degree of nonlocality of order $\alpha^{\textrm{b}}(y_0)$. The interaction within the layers shifts the neutral axis from $y=0$ (geometric axis) to $\tilde{y}=0$ (physical axis). $h_c$ is the relative position between physical and geometric center axis such that $(\Tilde{x},\Tilde{y})=(x,\Tilde{y})=(x,y-h_c)$.}
    \label{fig: Overview}
\end{figure}

A schematic illustration of the 2D microstructure of the beam, along with the corresponding DO representation is provided in Fig.~\ref{fig: Overview}. As evident from Fig.~\ref{fig: Overview}(a), the 2D microstructure is realized by (artificially) stacking in the transverse direction (i.e. the $y$-axis) different 1D nonlocal layers (oriented in the $x$-direction). The strength of the nonlocal effect (essentially, the CO characterizing each layer) associated with each 1D layer (that is, along the $x$-axis) is assumed to be uniform. However, there exists a functional gradation in the strength of the nonlocal effects when moving from layer to layer along the $y$-axis. Note that the localization of nonlocal effect within each (axially-oriented) layer implies that there are no direct nonlocal interactions within individual elements present on different layers; in other terms, this situation represents an anisotropic nonlocal effect. The same scenario can be presented in more mathematical terms, observing that $\alpha^{\textrm{b}}_x \in (0,1]$ while, $\alpha^{\textrm{b}}_y = 1$. Henceforth, for brevity, $\alpha^{\textrm{b}}_x$ is denoted simply as $\alpha^{\textrm{b}}$ while the local value for $\alpha^{\textrm{b}}_y$ is directly embedded in all model evaluations. It is important to clarify that the 1D chains of spherical elements (that appear as lumped masses at the first glance) should not be specifically construed as, for example, atomic chains, molecular chains, porous networks, or even 1D granular solids \cite{rapaport2004art,gonzalez2012nonlocal,patnaik2022role}. These 1D chains are used to merely allow a visual (abstract) representation of the localization of the nonlocal behavior in the different material layers; the origin of the nonlocal effects within each layer can any of the several nonlocality-inducing sources highlighted previously \cite{bavzant2000size,shaat2020review}.

This study has a three-fold objective whose relevance to the broader problem of multiscale nonlocal models is presented here below.
\begin{enumerate}[label=O\arabic*.]
    \item \textit{Development of the reference solution}: consists in leveraging the 3D DO nonlocal elasticity framework developed in \cite{ding2022multiscale} to obtain the reference solution for the fully-resolved 2D multiscale elasticity problem. Broadly speaking, this process consists in tailoring the DO operators to capture the anisotropic and heterogeneous variation of nonlocality which (as discussed before) is intrinsic to the considered 2D microstructure. This development is essential for two different reasons. First, on a theoretical note, it enables a very transparent understanding of the role of the tensorial definition of the DO strength function in capturing the physical effect of structural heterogeneity within a physical and mathematically consistent nonlocal model. More specifically, we will show how different components of this strength function can be carefully crafted to enable resolution (that is, top-down) or reduction (that is, bottom-up) of different (associated) material scales. Further, in the context of model validation and assessment (in O3), the 2D development is very critical because it provides a direct estimate of the computational cost incurred (or, equivalently, saved) in generating a specific DO representation, while still guaranteeing an accurate reference model (since, the 2D model does not compromise the representation of the different nonlocal interactions).
    
    \item \textit{Development of the theoretical DO model}: consists in developing a 1D DO constitutive model for the multiscale beam by employing the kinematic displacement-field assumptions from the Timoshenko's first-order shear deformation beam theory within the fully 3D DO continuum framework. Note that the Timoshenko kinematic assumptions \cite{timoshenko1955strength}, which enforce a uniform variation of the transverse displacement along the thickness of the beam, are consistent with the previous assumption that restricted the direct nonlocal interactions between individual layer elements in the benchmark problem (that is, $\alpha^{\textrm{b}}_y = 1$). The 1D DO governing equations for the multiscale beam are derived in a strong form using variational principles. This development is critical for three reasons. First, analogously to \cite{ding2022multiscale}, this development demonstrates the DO-FC based multiscale modeling principle in direct action. The 1D DO model is a reduced-order representation of the generalized 2D elasticity model (in O1) that captures the response of the fully-resolved 2D heterogeneous microstructure of the beam. Second, this development is critical since existing well-posed nonlocal beam models can only capture either anisotropic nonlocal effects \cite{sumelka2016fractional} or heterogeneous nonlocal effects \cite{patnaik2021variable}, but not both simultaneously as required in the present benchmark problem. Finally, the inverse simulation of the 1D DO problem also reveals remarkable insights on the natural handshaking ability of the DO model.
    
    \item \textit{Model validation and assessment}: consists in using both the 2D nonlocal elasticity (O1) and the 1D Timoshenko beam model (O2) to numerically analyze the static response of nonlocal beams. We will use a direct comparison of the results predicted from the fully resolved 2D model and the 1D DO model to analyze the accuracy and computational efficiency of the DO model.
\end{enumerate}

The remainder of this paper is organized as follows. First, in \S\ref{sec: anisotropic nonlocal elasticity}, we develop a generalized theoretic framework of anisotropic nonlocal elasticity that can be leveraged to describe the fully-resolved multiscale nonlocal beam (see O1 in \S\ref{ssec: Problem_setup}). Next, we then develop the 1D DO nonlocal Timoshenko beam theory in \S\ref{sec: Timoshenko beam modeling} for the DO-FC based multiscale modeling of nonlocal beams (O2). In \S\ref{sec: numerical studies}, we perform numerical simulations that address the mechanical response, the multiscale nonlocal effect, the numerical convergence, and the computational complexity, in order to demonstrate the performance of the DO approach (O3). Finally, we present some general remarks and conclusions in \S\ref{sec: conclusion}.

\section{Distributed-order anisotropic nonlocal elasticity theory}
\label{sec: anisotropic nonlocal elasticity}
In this section, we develop the anisotropic nonlocal elasticity theory that will be leveraged to determine the response of the multiscale beam (shown in Fig.~\ref{fig: Overview}) with a fully-resolved 2D microstructure (see O1). In order to model the response of the multiscale beam, according to the benchmark problem discussion in \S\ref{ssec: Problem_setup}, we develop a generalized nonlocal elasticity theory that allows capturing: 1) anisotropic nonlocality with nonlocal effects localized only in layers along the axial $x$-direction, and 2) the functional gradation of the strength of the axial nonlocal effects along the transverse $y$-direction. For this purpose, we leverage distributed-order operators and develop a generalized elasticity framework which will be henceforth referred to as the distributed-order anisotropic nonlocal elasticity theory (DO-ANET). We will also use the theoretical formulation to highlight the ability of the generalized DO theory in capturing both the anisotropic and heterogeneous nonlocal effects via a tailoring of the strength-function tensor.

\subsection{Theoretical framework}\label{ssec: theoretical_framework}
We first develop the mathematical and theoretical framework for DO-ANET that can model 2D nonlocal elastic beams. In analogy with the methodology presented in~\cite{ding2022multiscale}, based on the assumption of strain-driven nonlocality, the strain-displacement relationship remains unaltered and does not involve any fractional order operator. Hence, the kinematics relation under 2D plane strain assumptions is expressed in its classical form:
\begin{equation}\label{eq: varepsilon}
\bm{\varepsilon}=\frac{1}{2}\left(\bm{\nabla}\bm{u}(\bm{x})+\bm{\nabla}^\mathrm{T}\bm{u}(\bm{x})\right)
\end{equation}
where $\bm{\varepsilon}$ is the strain tensor, $\bm{u}$ is the displacement field, and $\bm{x}$ is the position in 2D Cartesian coordinates. Note that, in this approach, the kinematic formulation has a classical local description, while the constitutive relation is nonlocal and defined via DO fractional operators. The stress field can be expressed by making use of the constitutive relation in~\cite{ding2022multiscale} and takes the following form (provided in either tensor or indicial notation):
\begin{equation}\label{eq: sigma-DO-NET}
\begin{aligned}
    \bm{\sigma}&= \left(\prescript{\textrm{R-RL}}{a_1}{}\mathcal{I}_{b_1}^{1-\alpha,\bm{\kappa}(\alpha)} \circ \prescript{\textrm{R-RL}}{a_2}{}\mathcal{I}_{b_2}^{1-\alpha,\bm{\kappa}(\alpha)}\right) (\bm{C}:\bm{\epsilon})\\
    \sigma_{ij}&=\int_{0}^{1}\kappa_{ijkl}(\alpha) \odot  \left(\prescript{\textrm{R-RL}}{a_1}{}\bm{I}_{b_1}^{1-\alpha} \circ \prescript{\textrm{R-RL}}{a_2}{}\bm{I}_{b_2}^{1-\alpha}\right) \left(C_{ijkl}\epsilon_{kl}\right)\textrm{d}\alpha\\
\end{aligned}
\end{equation}
where $\odot$ represents the Hadamard product and $\circ$ represents a sequential operation ($f\circ(\cdot)\equiv f(g(\cdot))$. $\prescript{\textrm{R-RL}}{a_1}{}\mathcal{I}_{b_1}^{1-\alpha,\bm{\kappa}(\alpha)}(\cdot)$ and $\prescript{\textrm{R-RL}}{a_2}{}\mathcal{I}_{b_2}^{1-\alpha,\bm{\kappa}(\alpha)}(\cdot)$ shown in the tensor notation are DO Riesz type Riemann-Liouville fractional integrals (see detailed definitions of CO and DO fractional operators in \cite{ding2022multiscale} and in SM~\S1); $\prescript{\textrm{R-RL}}{a_1}{}\bm{I}_{b_1}^{1-\alpha,\bm{\kappa}(\alpha)}(\cdot)$ and $\prescript{\textrm{R-RL}}{a_2}{}\bm{I}_{b_2}^{1-\alpha,\bm{\kappa}(\alpha)}(\cdot)$ in the indicial notation are the corresponding CO Riesz-type Riemann-Liouville fractional integrals; $[a_1,b_1]$ and $[a_2,b_2]$ defined in both the DO and CO integrals are the nonlocal intervals in the two directions; $\bm{C}=C_{ijkl}$ (subscripts $\{i,j,k,l\} \in \{1,2\}$ represent the two orthonormal axes in 2D Cartesian space) is the material stiffness tensor; $\bm{\kappa}(\alpha)=\kappa_{ijkl}(\alpha)$ is the fourth-order strength-function tensor that satisfies the partition of unity condition:
\begin{equation}\label{eq: partition_of_unity}
    \int_{0}^{1}\kappa_{ijkl}(\alpha)\mathrm{d}\alpha = 1
\end{equation}
The order $\alpha$ is bound to vary within the interval $[0,1]$ to allow for a fractional order representation of the nonlocal stress. Note that similar to the stiffness tensor $\bm{C}$ that describes the material properties (either isotropic or anisotropic), the strength-function tensor $\bm{\kappa}(\alpha)$ is used to describe the nonlocal properties (either isotropic or anisotropic). In this regard, the two fourth-order tensors should share the same mathematical properties (such as major and minor symmetry, see \cite{ding2022multiscale}) and $\bm{\kappa}$ should follow the Hadamard product rule with $\bm{C}$.

While the use of a strength-function tensor allows capturing anisotropic nonlocality, the formulation still cannot address heterogeneous nonlocality in the $y$-direction, as shown in Fig.~\ref{fig: Overview}. To further generalize the constitutive relation and address both anisotropic and heterogeneously distributed nonlocal behavior, we extend the spatially-independent `strength-function tensor' by introducing distributed-variable-order fractional operators \cite{ding2021applications}. Specifically, by rendering the strength function dependent on both the order $\alpha$ and the spatial coordinates $\bm{x}\equiv(x,y)$, the DO constitutive relations in Eq.~(\ref{eq: sigma-DO-NET}) can be updated as:
\begin{equation}\label{eq: sigma-DVO-NET}
\begin{aligned}
    \bm{\sigma} &= \left(\prescript{\textrm{R-RL}}{a_1}{}\mathcal{I}_{b_1}^{1-\alpha,\bm{\kappa}(\alpha,\bm{x})} \circ \prescript{\textrm{R-RL}}{a_2}{}\mathcal{I}_{b_2}^{1-\alpha,\bm{\kappa}(\alpha,\bm{x})}\right)(\bm{C}:\bm{\epsilon})\\
    \sigma_{ij} &= \int_{0}^{1}\kappa_{ijkl}(\alpha,x,y) \odot \left(\prescript{\textrm{R-RL}}{a_1}{}\bm{I}_{b_1}^{1-\alpha} \circ \prescript{\textrm{R-RL}}{a_2}{}\bm{I}_{b_2}^{1-\alpha}\right) \left(C_{ijkl}\epsilon_{kl}\right)\textrm{d}\alpha\\
\end{aligned}
\end{equation}
where also the updated strength-function tensor should also satisfy the partition of unity (see Eq.~(\ref{eq: partition_of_unity})). Note that, different from the classical definition of variable-order (VO) fractional operators where the order $\alpha=\alpha(\bm{x})$ is spatially-dependent, Eq.~(\ref{eq: sigma-DVO-NET}) leverages the spatially variable strength-function tensor $\bm{\kappa}=\bm{\kappa}(\alpha, \bm{x})$ to achieve a spatially-dependent feature. Recall that in DO fractional operators, $\bm{\kappa}(\alpha)$ is exploited to capture the strength (or the portion) of each specific order $\alpha$ \cite{ding2022multiscale}. By enforcing $\bm{\kappa}$ to be spatially dependent, the strength of each order $\alpha$ in the spatially-dependent DO operators can be tuned accordingly to capture the variation of nonlocal order in space (for example, the heterogeneous distribution of order $\alpha^{\mathrm{b}}(y)$ in the $y$-direction shown in Fig.~\ref{fig: Overview}). It immediately follows that the DO formulation is naturally equipped to account for the effect of structural heterogeneity (that manifest as nonlocal effects) directly within a strain-driven integral constitutive framework; this approach bears similarities to the seminal proposition from Eringen \cite{eringen1972nonlocal} but without using an anisotropic elasticity tensor. The result is remarkable because, as noted in \cite{batra159misuse}, several existing nonlocal approaches that model heterogeneous structures via Eringen's integral formulation, directly violate the material isotropy assumption in \cite{eringen1972nonlocal} by using an anisotropic elasticity tensor.

Based on the formulations for both strain and stress fields, the governing equations and associated boundary conditions for the anisotropic nonlocal elastic solid can be derived using the Hamilton's principle. Following the procedure provided in \cite{ding2022multiscale}, we obtain the strong form of the governing equations:
\begin{equation}\label{eq: governing equation}
\bm{\nabla}\cdot\bm{\sigma}+\rho\bm{b}={\rho}\Ddot{\bm{u}}
\end{equation}
with displacement and traction boundary conditions:
\begin{equation}\label{eq: BCs}
    \bm{u}=\bm{u}_0,\quad \bm{\sigma}\cdot\bm{n}-\bm{T}=0
\end{equation}
where $\bm{b}$ indicates body forces, $\bm{u}_0$ is the prescribed displacement, $\bm{n}$ is the normal vector to the surface, and $\bm{T}$ is the surface traction.Note that, since we introduced the nonlocal behavior via fractional constitutive relations (as opposed to fractional kinematics approach~\cite{patnaik2020generalized,patnaik2021towards}), all the nonlocal information is isolated within the nonlocal stress $\bm{\sigma}$ such that the form of both the governing equations and the boundary conditions is apparently independent from the DO formulation and remain identical to the their counterparts in local linear elastic theory. Equations~(\ref{eq: governing equation},\ref{eq: BCs}) complete the theoretical framework for DO-ANET that will be used in the following sections to model the multiscale nonlocal beams.

\subsection{2D DO formulation of anisotropic nonlocal elastic beams}
We leverage the DO-ANET to model the behavior of the nonlocal elastic beams presented in Fig.~\ref{fig: Overview} and subject to nonlocal anisotropy (i.e. the nonlocal interactions act only in the $x$-direction) and nonlocal heterogeneity (i.e. the order of nonlocality is distributed heterogeneously in the $y$-direction). As we previously demonstrated, by introducing the DO operator and the strength-function tensor, the two nonlocal elastic characteristics can be captured simultaneously via the proposed DO-ANET in \S\ref{ssec: theoretical_framework}. Following the theoretical framework, we first define the spatially-dependent strength-function tensor, at a given point $(x,y)$:
\begin{equation}\label{eq: kappa}
\kappa_{\beta\gamma}(\alpha,y)=
\begin{bmatrix}
\kappa_{11}(\alpha,y) & \delta(\alpha-1) & \delta(\alpha-1)\\
\delta(\alpha-1) & \delta(\alpha-1) & \delta(\alpha-1) \\
\delta(\alpha-1) & \delta(\alpha-1) & \delta(\alpha-1) \\
\end{bmatrix}
\end{equation}
where $\delta(\cdot)$ is the Dirac delta function. For the sake of simplicity and clarity, we use the Voigt form $\kappa_{\beta\gamma}$ ($\{\beta,\gamma\} \in \{1,2,3\}$) of the strength-function tensor to explicitly show the mathematical expression of each component. The same rule also applies to stiffness, strain, and stress tensors in the following derivation. Note that to capture the nonlocal anisotropy (which shows nonlocality in the $x$-direction only), we leave the first component $\kappa_{11}$ to be undetermined (so to be defined in accordance to specific types of nonlocality) while define the remaining components in $\kappa_{\beta\gamma}$ all as $\delta(\alpha-1)$ to represent integer order derivatives (i.e. local elastic components). Also note that to capture transverse heterogeneity in the nonlocal behavior (represented by $\alpha^{\textrm{b}}(y)$), we define the component $\kappa_{11}=\kappa_{11}(\alpha,y)$ to be an explicit function of the discrete position in the $y$-direction; $\kappa_{11}$ is not considered as a function of the longitudinal position $x$ simply because the nonlocal order $\alpha$ remains constant (by initial assumption) in the $x$-direction. By leveraging the Dirac-delta function, the strength-function $\kappa_{11}$ can be defined by:
\begin{equation}\label{eq: kappa_11}
    \kappa_{11}(\alpha,y)=\delta(\alpha-\alpha^{\mathrm{b}}(y))
\end{equation}
to capture the heterogeneous nonlocal behavior in the $y$-direction as illustrated in Fig.~\ref{fig: Overview}.

Following the above definitions, the stress tensor for 2D (plane strain) beam problems can be formulated via DO nonlocal elasticity. Specifically, we consider a beam with homogeneous elastic material properties. The stiffness tensor $C_{ijkl}$ for 2D isotropic elasticity in Voigt notation is given by:
\begin{equation}\label{eq: C}
    C_{\beta\gamma}=
    \begin{bmatrix}
    2\mu+\lambda & \lambda & 0\\
    \lambda & 2\mu+\lambda & 0\\
    0 & 0 & 2\mu\\
    \end{bmatrix}
\end{equation}
Substituting $\kappa_{\beta\gamma}$ and $C_{\beta\gamma}$ into the DO constitutive relation in Eq.~(\ref{eq: sigma-DO-NET}), we obtain the stress in its tensorial and Voigt notation form:
\begin{equation}\label{eq: sigma-anisotropic}
\begin{aligned}
    \bm{\sigma}&=\int_{0}^{1}\bm{\kappa}(\alpha,y)\prescript{\textrm{R-RL}}{0}{}\bm{I}_{L}^{1-\alpha}\left(\bm{C}:\bm{\varepsilon}\right)\mathrm{d}\alpha \\
    \sigma_{\beta}&=\left[(2\mu+\lambda)\int_{0}^{1}\kappa_{11}(\alpha,y)\cdot\prescript{\textrm{R-RL}}{0}{}\bm{I}_{L}^{1-\alpha}\left(\varepsilon_{1}\right)\mathrm{d}\alpha+\lambda\varepsilon_{2},~\lambda\varepsilon_{1}+(2\mu+\lambda)\varepsilon_{2},~2\mu\varepsilon_{3}\right]^\mathrm{T} \\
\end{aligned}
\end{equation}
where $\prescript{\textrm{R-RL}}{0}{}\bm{I}_{L}^{1-\alpha}(\cdot)$ is the Riesz type Riemann-Liouville fractional integral operator in the $x$-direction with the nonlocal horizon defined as $[0,L]$ (which is identical to the configuration in Fig.~\ref{fig: Overview}) and $\varepsilon_{\gamma}=[\varepsilon_1,\varepsilon_2,\varepsilon_3]^T$ is the Voigt form of strain tensor $\bm{\varepsilon}$. Further substituting Eqs.~(\ref{eq: kappa_11},\ref{eq: C},\ref{eq: sigma-anisotropic}) into Eqs.~(\ref{eq: governing equation},\ref{eq: BCs}) and using the 2D Cartesian coordinates $x$-$y$ shown in Fig.~\ref{fig: Overview}, we obtain the explicit form of the governing equations for 2D nonlocal plane strain beam problems:
\begin{subequations}\label{eq: governing_eqn_beam}
\begin{align}
    \rho \Ddot{u}_x & = (2\mu+\lambda)\left[\prescript{\textrm{R-C}}{0}{}\bm{D}_{L}^{\alpha^{\textrm{b}}(y)}u_{x}\right]_{,x} + (\mu+\lambda)u_{y,xy} + {\mu}u_{x,yy} + f_x \\
    \rho \Ddot{u}_y & = (2\mu+\lambda)u_{y,yy} + (\mu+\lambda)u_{x,xy} + {\mu}u_{y,xx} + f_y
\end{align}
\end{subequations}
where (expressed in displacement $u_x$ and $u_y$ that are defined in the 2D Cartesian coordinates shown in Fig.~\ref{fig: Overview}); $\prescript{\textrm{R-C}}{0}{}\bm{D}_{L}^{\alpha^{\textrm{b}}(y)}(\cdot)$ is the Riesz-Caputo fractional derivative in the $x$-direction which is closely related to the Riemann-Liouville definition of fractional operators (see the relation between different definitions of fractional operators in SM~\S1). We highlight that, by taking the definition of strength-function tensor $\bm{\kappa}$ given in Eq.~(\ref{eq: kappa}), only the governing equation in the $x$-direction involves fractional operators and the remaining governing equation in the $y$-direction is identical to its local form. It is also critical to note that, although both the strength-function tensor and the fractional derivative operators are spatially dependent in the $y$-direction, they should be distinguished from the variable-order (VO) fractional operators~\cite{patnaik2021variable}. Recall that in the VO formulation, fractional operators apply in the same direction the fractional order depends explicitly on. For example, the fractional order must be a function of $x$ if the VO derivative is taken in the $x$-direction; as a contrast, the fractional derivative formulated in Eq.~(\ref{eq: governing_eqn_beam}a) is taken in the $x$-direction while the order $\alpha^\mathrm{b}$ depends on $y$, not $x$. In general, Eq.~(\ref{eq: governing_eqn_beam}) can be used to describe the mechanical behavior of 2D beams with nonlocal anisotropy, as well as nonlocal heterogeneity.

\section{Distributed-order nonlocal Timoshenko beam model}
\label{sec: Timoshenko beam modeling}
The previous section presented the theoretical formulation of the DO-ANET. While the theory allows describing the multiscale nonlocal behavior of beams, the solution requires full 2D discretization, hence leading to high computational cost. To address this shortcoming, in the following we use the DO-ANET presented above to develop a 1D DO nonlocal Timoshenko beam formulation (see O2 in \S\ref{ssec: Problem_setup}). Specifically, by leveraging the multiscale nonlocal characteristic of DO operators, we develop a DO nonlocal Timoshenko beam model (DO-NTBM) that can capture both 1) the overall uniaxial nonlocal response, and 2) the heterogeneous distribution of nonlocal order in the $y$-direction.

\subsection{Derivation of DO-NTBM and governing equations}\label{ssec: Timoshenko governing equation}
We develop the DO-NTBM from the framework of anisotropic nonlocal elasticity in \S\ref{sec: anisotropic nonlocal elasticity}. Due to the heterogeneous distribution of nonlocality, a traditional definition of the displacement field (as typically used in beam theories) is inadequate to capture the disturbance resulting from the varying nonlocal effects. Indeed, according to classical beam theories, the neutral axis of a homogeneous beam undergoing pure bending lies on the centroid of the cross section and can be easily obtained by evaluating the first moment of area~\cite{timoshenko1955strength}. However, since in this study we consider a beam with heterogeneously distributed nonlocality, the position of neutral axis can be altered by the heterogeneous distribution and eventually does not coincide with the geometric center line~\cite{timoshenko1955strength}. To account for this aspect, we consider defining two distinct sets of axes in the axial direction: 1) geometric center axis that connects the geometric centers of the cross sections, and 2) physical center axis that serves as the real neutral axis of mechanical properties. Figure~\ref{fig: Overview}(b) shows the detailed configurations of the two sets of axes ($x$-$y$ and $x$-$\Tilde{y}$) and their corresponding coordinate systems. According to the assumptions for the Timoshenko beam formulation~\cite{timoshenko1955strength}, the displacement field at a given point $(x,\Tilde{y})$ in physical coordinates (or equivalently, $(x,y)$ in geometric coordinates) under transverse loading condition is given by:
\begin{equation}\label{eq: displacement-Timoshenko}
\begin{aligned}
    u_x(x,\Tilde{y})&=-\Tilde{y}{\phi}(x)\\
    u_y(x,\Tilde{y})&=w(x)\\
\end{aligned}
\end{equation}
where $\Tilde{y}=y-h_c$ is the transverse position in physical coordinates, $w(x)$ is the transverse component of the displacement field, and ${\phi}(x)$ is the angle of rotation of the normal to the physical center axis. Note that since $x$ and $\Tilde{x}$ axes are equivalent, in the following we use $x$ as the axial position of material points; also, we use the symbol $\Tilde{\square}$ to distinguish quantities that are defined exclusively in physical coordinates.

To obtain the strain field, we employ the same local kinematics formulated in \S\ref{sec: anisotropic nonlocal elasticity}. Specifically, since the kinematics in Eq.~(\ref{eq: varepsilon}) does not involve nonlocal formulation, the strain field for the nonlocal Timoshenko beam can be simply obtained by substituting the displacement field Eq.~(\ref{eq: displacement-Timoshenko}) into Eq.~(\ref{eq: varepsilon}):
\begin{equation}\label{eq: varepsilon-Timoshenko}
\begin{aligned}
    \Tilde{\varepsilon}_{xx}(x,\Tilde{y}) &= -\Tilde{y} \phi^{\prime}(x) \\
    \Tilde{\varepsilon}_{x\Tilde{y}}(x,\Tilde{y}) &= \Tilde{\varepsilon}_{x\Tilde{y}}(x)=\frac{1}{2}\left[w^{\prime}(x)-{\phi}(x)\right] \\
\end{aligned}
\end{equation}
where $\Tilde{\varepsilon}_{xx}$ and $\Tilde{\varepsilon}_{x\Tilde{y}}$ are the axial and shear strains, respectively; $(\cdot)^{\prime}=\mathrm{d}(\cdot)/\mathrm{d}x$ denotes the first order spatial derivative in the $x$-direction. In analogy with the classical Timoshenko beam theory, only the axial strain $\Tilde{\varepsilon}_{xx}$ explicitly depends on the transverse position $\Tilde{y}$ while the shear strain $\Tilde{\varepsilon}_{xy}$ remains constant through out the transverse direction.

To formulate the stress field for the nonlocal Timoshenko beam theory, we use the nonlocal constitutive relations. Note that, according to the Timoshenko beam assumptions, only the axial stress $\sigma_{xx}$ and the shear stress $\sigma_{x\Tilde{y}}$ are required to describe the beam's stress state. The stress state in our formulation will capture both the anisotropic nonlocal interactions and also the shear deformation associated with the heterogeneous distribution of nonlocality. To derive these stress components, we leverage the prior results from the DO-ANET framework. Specifically, by adopting the anisotropic constitutive relation in Eq.~(\ref{eq: sigma-anisotropic}) and the strength-function $\kappa_{11}(\alpha,y)$ in Eq.~(\ref{eq: kappa_11}), we obtain:
\begin{equation}\label{eq: sigma-Timoshenko}
\begin{aligned}
    \Tilde{\sigma}_{xx}(x,\Tilde{y}) &= -\Tilde{y}E\left[\prescript{\textrm{R-C}}{0}{}\bm{D}_{L}^{\Tilde{\alpha}^{\textrm{b}}(\Tilde{y})}\phi(x)\right] \\
    \Tilde{\sigma}_{x\Tilde{y}}(x,\Tilde{y}) &= \Tilde{\sigma}_{x\Tilde{y}}(x) = \mu\chi\left[w^{\prime}(x)-\phi(x)\right] \\
\end{aligned}
\end{equation}
where $\chi$ is the Timoshenko shear coefficient used to compensate the non-uniform distribution of actual shear stress over the beam's cross section. $\Tilde{\alpha}^{\textrm{b}}(\Tilde{y})=\alpha^{\textrm{b}}(\Tilde{y}+h_c)$ is the nonlocal order defined in physical coordinates. Note that unlike classical Timoshenko theory that treats axial stress as a homogeneously distributed quantity, the current $\Tilde{\sigma}_{xx}$ is now a function of $\Tilde{y}$ (see the fractional order $\Tilde{\alpha}^{\textrm{b}}(\Tilde{y})$ in Eq.~(\ref{eq: sigma-Timoshenko})) due to the heterogeneous distribution of nonlocality in the transverse direction. Similar to the $\hat{\varepsilon}_{x\Tilde{y}}$, the shear stress $\hat{\sigma}_{x\Tilde{y}}$ also remains constant in the transverse direction.

Based on the above definitions of the strain and stress fields, we now derive the strong form of the governing equations and of the associated boundary conditions by using variational principles. Considering the deformation energy and the work done by external forces, the Hamiltonian of the nonlocal Timoshenko beam is given by:
\begin{equation}\label{eq: Hamiltonian}
    \mathcal{H}=\int_{t_0}^{t_1}\left[\mathbb{U}-\int_{0}^{L}q(x)w(x)\mathrm{d}x\right]\mathrm{d}t
\end{equation}
where $q(x)$ is the transverse external force per unit length applied on the nonlocal Timoshenko beam. $\mathbb{U}$ is the total deformation energy defined as:
\begin{equation}
    \mathbb{U}=\frac{1}{2}\int_{\Omega}\left(\Tilde{\sigma}_{xx}\Tilde{\varepsilon}_{xx}+\Tilde{\sigma}_{x\Tilde{y}}\Tilde{\varepsilon}_{x\Tilde{y}}\right)\mathrm{d}\Omega
\end{equation}
where $\Omega$ is the volume of the beam. By applying the Hamilton's principle and standard rules of variational calculus to Eq.~(\ref{eq: Hamiltonian}), the strong form of the governing equations is found to be:
\begin{equation}\label{eq: Timoshenko equation-1}
\begin{aligned}
    M^{\prime}(x)-Q(x)&=0\\
    Q^{\prime}(x)+q(x)&=0\\
\end{aligned}
\end{equation}
with three types of possible boundary conditions:
\begin{equation}\label{eq: beam boundary conditions}
\begin{aligned}
    \textrm{Clamped end:}& \quad w(x)=0 \quad \textrm{and} \quad \phi^{\prime}(x)=0 \quad &\forall x \in \{0,L\} \\
    \textrm{Simply supported end:}& \quad w(x)=0 \quad \textrm{and} \quad M(x)=M_0(x) \quad &\forall x \in \{0,L\} \\
    \textrm{Free end:}& \quad Q(x)=Q_0(x) \quad \textrm{and} \quad M(x)=M_0(x) \quad &\forall x \in \{0,L\}\\
\end{aligned}
\end{equation}
where $M(x)$ is the bending moment and $Q(x)$ is the shear force. Integrating over the cross section yields:
\begin{equation}\label{eq: M-Q}
\begin{aligned}
    M(x)&=\int_{-\frac{h}{2}-h_c}^{\frac{h}{2}-h_c} \Tilde{y}\Tilde{\sigma}_{xx}\mathrm{d}\Tilde{y}
    =
    -E\int_{-\frac{h}{2}-h_c}^{\frac{h}{2}-h_c} {\Tilde{y}}^2\left[\prescript{\textrm{R-C}}{0}{}\bm{D}_{L}^{\Tilde{\alpha}^{\textrm{b}}(\Tilde{y})}\phi(x)\right]\mathrm{d}\Tilde{y} \\
    Q(x)&=\int_{-\frac{h}{2}-h_c}^{\frac{h}{2}-h_c}\Tilde{\sigma}_{x\Tilde{y}}\mathrm{d}\Tilde{y}
    =h\mu\chi[w^{\prime}(x)-\phi(x)]
\end{aligned}
\end{equation}
Where the upper and lower bounds of integration are the coordinates of the upper and lower surfaces of the beam in the physical coordinates. Note that, the formulation of the bending moment $M(x)$ involves the integration of fractional derivatives with different order (due to the nonlocal heterogeneity). Since a 2D beam is considered in this study, we ignore the variation of the beam width (perpendicular to both the longitudinal and transverse directions) in the calculation of $M(x)$ and $Q(x)$. To satisfy the dimension consistency, we consider a unit width $1[\mathrm{m}]$ throughout the whole beam. The integro-differential character of $M(x)$ suggests that we can define a DO derivative:
\begin{equation}
    \prescript{\textrm{R-C}}{0}{}\mathcal{D}_{L}^{\Tilde{\alpha}^{\textrm{b}}(\Tilde{y})}f(x) = \int_{-\frac{h}{2}-h_c}^{\frac{h}{2}-h_c}\left[\Tilde{\kappa}(\Tilde{y})\prescript{\textrm{R-C}}{0}{}\bm{D}_{L}^{\Tilde{\alpha}^{\textrm{b}}(\Tilde{y})}f(x)\right]\mathrm{d}\Tilde{y}\\
\end{equation}
such that the information on the nonlocal heterogeneity is entirely embedded within a single operator. In order to ensure 1) the reduction of the 1D DO beam model to the classical Timoshenko beam model for an isotropic local beam, and 2) the dimensional consistency, the strength-function $\Tilde{\kappa}(\Tilde{y})$ is defined as:
\begin{equation}
\label{eq: strength_function}
    \Tilde{\kappa}(\Tilde{y})=\frac{\Tilde{y}^2} {\displaystyle\int_{-\frac{h}{2}-h_c}^{\frac{h}{2}-h_c}\Tilde{y}^2\mathrm{d}\Tilde{y}} = \frac{\Tilde{y}^2}{\Tilde{I}},\quad \int_{-\frac{h}{2}-h_c}^{\frac{h}{2}-h_c}\Tilde{\kappa}(\Tilde{y})\mathrm{d}\Tilde{y} = 1
\end{equation}
such that the bending moment $M(x)$ in Eq.~(\ref{eq: M-Q}) can be written in the form of a DO derivative:
\begin{equation}\label{eq: M-2}
    M(x)=-E\Tilde{I}\left[\prescript{\textrm{R-C}}{0}{}\mathcal{D}_{L}^{\Tilde{\alpha}^{\textrm{b}}(\Tilde{y})}\phi(x)\right]
\end{equation}
where $\Tilde{I}=\left(\frac{h^3}{12}+hh_c^2\right)\cdot 1[\textrm{m}]$ is the second moment of area for the cross section, which is expressed in units of $[\textrm{m}^4]$. 
By substituting Eq.~(\ref{eq: M-2}) back into Eq.~(\ref{eq: Timoshenko equation-1}), we obtain the DO fractional governing equations for the nonlocal Timoshenko beam:
\begin{equation}\label{eq: Timoshenko equation-2}
\begin{aligned}
    E\Tilde{I}\left[\prescript{\textrm{R-C}}{0}{}\mathcal{D}_{L}^{\Tilde{\alpha}^{\textrm{b}}(\Tilde{y})}\phi(x)\right]^{\prime\prime}&=q(x)\\
    \phi(x)-\frac{E\Tilde{I}}{h\mu\chi}\left[\prescript{\textrm{R-C}}{0}{}\mathcal{D}_{L}^{\Tilde{\alpha}^{\textrm{b}}(\Tilde{y})}\phi(x)\right]^{\prime}&=w^{\prime}(x)\\
\end{aligned}
\end{equation}
where the single and double prime indicate the the first and second integer order spatial derivative, respectively. Combining Eq.~(\ref{eq: Timoshenko equation-2}) with the boundary conditions in Eq.~(\ref{eq: beam boundary conditions}), we complete the derivation of the uniaxial nonlocal Timoshenko beam theory using DO fractional derivatives. We highlight that by combining the heterogeneous nonlocal behavior within the DO derivative, the transverse direction in the original 2D nonlocal beams can be \enquote{collapsed}, hence transforming the original 2D nonlocal model into the 1D DO-NTBM; this procedure effectively results in a model order reduction approach. Similar to classical beam theories, by ignoring the shear stress in the second equation of Eq.~(\ref{eq: Timoshenko equation-2}), we recover a DO Euler-Bernoulli beam equation which ignores the shear effects over the cross section.

Note that the derivation of the strength function in Eq.~(\ref{eq: strength_function}) provides an immediate inverse approach to determine the strength function of the overall model. This inverse strategy merely consists in enforcing a reduction of the DO Timoshenko model to the classical Timoshenko model for an isotropic microstructure. This outcome is remarkable since it demonstrates that the DO model is parsimonious in nature. More specifically, the DO nonlocal approach enables a successful representation of the multiscale nonlocal phenomenon with the least possible predictor variables (here only the strength-function) \cite{regenwetter2018heterogeneity}.

\subsection{Derivation of auxiliary equations}\label{ssec: auxiliary equations}
Note that unlike the classical Timoshenko beam theory, the present formulation of DO-NTBM includes two extra parameters (i.e. $h_c$ and $\chi$) that capture the effect of the heterogeneous nonlocal order distribution over the cross section. The system of governing equations in DO-NTBM (Eq.~(\ref{eq: Timoshenko equation-2})) must be complemented by two independent auxiliary equations in order to determine these two additional parameters.

First, we derive the auxiliary equation for $h_c$ which determines the position of the physical center axis. Knowing that, for beams under pure transverse loading, the axial force over the cross section must vanish:
\begin{equation}\label{eq: N}
    N(x)=\int_{-\frac{h}{2}-h_c}^{\frac{h}{2}-h_c}\Tilde{\sigma}_{xx}\mathrm{d}\Tilde{y}
    =-E\int_{-\frac{h}{2}-h_c}^{\frac{h}{2}-h_c}\Tilde{y}\left[\prescript{\textrm{R-C}}{0}{}\bm{D}_{L}^{\Tilde{\alpha}^{\textrm{b}}(\Tilde{y})}\phi(x)\right]\mathrm{d}\Tilde{y}=0
\end{equation}
where $N(x)$ is the axial stress resultant at a given axial position $x$. Eq.~(\ref{eq: N}) can be manipulated to obtain $h_c$:
\begin{equation}\label{eq: h_c-1}
    h_c(x) = \frac{\displaystyle\int_{-\frac{h}{2}}^{\frac{h}{2}}y\left[\prescript{\textrm{R-C}}{0}{}\bm{D}_{L}^{\alpha^{\textrm{b}}(y)}\phi(x)\right]\mathrm{d}y}{\displaystyle\int_{-\frac{h}{2}}^{\frac{h}{2}}\left[\prescript{\textrm{R-C}}{0}{}\bm{D}_{L}^{\alpha^{\textrm{b}}(y)}\phi(x)\right]\mathrm{d}y}
\end{equation}
Note that, since the physical and geometric center axis are parallel, Eq.~(\ref{eq: h_c-1}) should hold at any position within the beam. This consideration allows recasting Eq.~(\ref{eq: h_c-1}) in the following form:
\begin{equation}\label{eq: h_c-2}
    h_c - \frac{1}{L}\displaystyle\int_{0}^{L} \left[\frac{\displaystyle\int_{-\frac{h}{2}}^{\frac{h}{2}}y\left[\prescript{\textrm{R-C}}{0}{}\bm{D}_{L}^{\alpha^{\textrm{b}}(y)}\phi(x)\right]\mathrm{d}y} {\displaystyle\int_{-\frac{h}{2}}^{\frac{h}{2}}\left[\prescript{\textrm{R-C}}{0}{}\bm{D}_{L}^{\alpha^{\textrm{b}}(y)}\phi(x)\right]\mathrm{d}y}\right] \mathrm{d}x = 0\\
\end{equation}
where $h_c$ becomes a constant for any $x \in [0,L]$. It is seen that, for nonlocal beams with symmetric distributed-order functions $\alpha^{\textrm{b}}(y)$ (that is $\alpha^{\textrm{b}}(y)=\alpha^{\textrm{b}}(-y)$), the integrand $y\left[\prescript{\textrm{R-C}}{0}{}\bm{D}_{L}^{\alpha^{\textrm{b}}(y)}\phi(x)\right]$ at the numerator in Eq.~(\ref{eq: h_c-2}) is an odd function with respect to $y$ and thus $h_c=0$, which means that the physical and the geometric center axes coincide. This should not be surprising given that the offset between the two sets of axes was due to the heterogeneity in nonlocality. On the other hand, for beams with asymmetric distributed-order functions (see the example of asymmetric distributed-order in Fig.~\ref{fig: Overview}(b)), the integral in the numerator is generally nonzero and thus it leads to the two inequivalent definitions between physical and geometric center axis. While classical Timoshenko beam model cannot capture this effect, the DO-NTBM formulation can effectively keep it into account via the auxiliary parameter $h_c$.

As for the determination of the shear coefficient $\chi$, we introduce a method based on shear strain energy matching~\cite{adamek2015analytical}. Recall that according to the Timoshenko beam assumptions, the real heterogeneous distribution of shear strain and stress over the cross section are substituted by homogeneous values (see Eq.~(\ref{eq: varepsilon-Timoshenko}) for strain and Eq.~(\ref{eq: sigma-Timoshenko}) for stress). In order to guarantee that the real heterogeneous distribution of shear stress is properly mapped into a nominal uniform distribution of shear stress, the method based on shear strain energy matching prescribes that the total shear strain energy stored in the beam should remain unaltered in the two representations.

To establish the equality of the total shear strain energy, we first derive the expression of the real distribution of the heterogeneous shear stress. Consider the force balance in the axial direction at point $(x,y^*)$ in the geometric coordinates with respect to an infinitesimal volume element:
\begin{equation}
\begin{aligned}
    \sigma_{xy}(x,y^*)\mathrm{d}x &=\int_{y^*}^{\frac{h}{2}}\Tilde{\sigma}_{xx}(x+\mathrm{d}x)\mathrm{d}y-\int_{y^*}^{\frac{h}{2}}\Tilde{\sigma}_{xx}(x)\mathrm{d}y =\int_{y^*}^{\frac{h}{2}}\Tilde{\sigma}_{xx}^{\prime}(x)\mathrm{d}x\mathrm{d}y \\
    \sigma_{xy}(x,y^*) &= \int_{y^*}^{\frac{h}{2}}\Tilde{\sigma}_{xx}^{\prime}(x)\mathrm{d}y=\int_{y^*}^{\frac{h}{2}}-yE\left[\prescript{\textrm{R-RL}}{0}{}\bm{D}_{L}^{\alpha^{\textrm{b}}(y)}\phi(x)\right]^{\prime}\mathrm{d}y \\
\end{aligned}
\end{equation}
where $\sigma_{xy}$ is the real distribution of shear stress that is heterogeneous over the beam's cross section. Note that we use $y^*$ to distinguish the lower bound of integration from the differential term $\mathrm{d}y$ in the integral formulation. In contrast to $\Tilde{\sigma}_{x\Tilde{y}}$ in Eq.~(\ref{eq: sigma-Timoshenko}), here we use subscript $y$ (not $\Tilde{y}$) to indicate that $\sigma_{xy}$ is defined in the geometric coordinates (same for $\Tilde{\sigma}_{xy}$ used below). Assuming that $\chi$ is a spatially independent constant, the equality of shear strain energy can be formulated as:
\begin{equation}\label{eq: chi-1}
    \Tilde{U}_{xy}=\frac{1}{2}\int_{0}^{L}\int_{-\frac{h}{2}}^{\frac{h}{2}}\frac{\Tilde{\sigma}_{xy}^2(x)}{\chi\mu}\mathrm{d}y\mathrm{d}x=U_{xy}=\frac{1}{2}\int_{0}^{L}\int_{-\frac{h}{2}}^{\frac{h}{2}}\frac{\sigma_{xy}^2(x,y)}{\mu}\mathrm{d}y\mathrm{d}x
\end{equation}
where $\Tilde{\sigma}_{xy}(x)=\Tilde{\sigma}_{x\Tilde{y}}(x)$ is the shear stress (which remains constant in the transverse direction) in geometric coordinates. To demonstrate the difference between the two types of shear stress, in the above formulation we use variable $x$ only to 
highlight the uniform distribution of $\sigma_{xy}(x)$ and in contrast, we use both $x$ and $y$ in $\Tilde{\sigma}_{xy}(x,y)$ to show its explicit dependency on $y$. Reorganizing the formulation, we derive the following equation that gives the explicit formulation of $\chi$:
\begin{equation}\label{eq: chi-2}
    \chi-\frac{\displaystyle\int_{0}^{L}\int_{-\frac{h}{2}}^{\frac{h}{2}}{\Tilde{\sigma}_{xy}^2(x)}\mathrm{d}y\mathrm{d}x}{\displaystyle\int_{0}^{L}\int_{-\frac{h}{2}}^{\frac{h}{2}}{\sigma_{xy}^2(x,y)}\mathrm{d}y\mathrm{d}x} = 0
\end{equation}
Equations~(\ref{eq: h_c-2},\ref{eq: chi-2}) provide the two additional equations that allow the solution of the DO Timoshenko beam formulation (see Eq.~(\ref{eq: Timoshenko equation-2})). Recall that the two extra parameters $h_c$ and $\chi$ are introduced to better capture the heterogeneous nature of nonlocality and to further demonstrate the DO operator's capability to model multiscale nonlocal effects. We emphasize that, while DO operators have been exploited in both the 2D DO-ANET and the 1D DO-NTBM, they play different roles and capture different physical mechanisms: in DO-ANET, the fourth-order kernel-function tensor is defined, for mathematical purpose, to develop the theoretic framework of anisotropic nonlocal elasticity; in DO-NTBM, the multiscale nature of DO operator is exploited to collect all the heterogeneous nonlocal information and hence to further capture the multiscale nonlocal elastic behavior. Based on the two previously derived modeling approaches, in the following sections we will perform multiscale numerical studies to simulate the uniaxial nonlocal beam problem.

\section{Numerical analyses of multiscale nonlocal beams}\label{sec: numerical studies}
In this section, we will perform comprehensive numerical studies using both the 2D DO-ANET and the 1D DO-NTBM (see O3 in \S\ref{ssec: Problem_setup}). More specifically, we numerically simulate the mechanical response of uniaxial nonlocal beams using both modeling approaches and perform a series of analyses to substantiate the effectiveness of the proposed multiscale modeling approach. Given the fundamental differences between the two models, simulation results that involve numerical accuracy, convergence, and computational cost are also different. Similar to classical (local) continuum theory whose solutions are generally treated as ground truth in the field of structural mechanics, in the following we will also take the results of DO-ANET as reference solutions for the multiscale nonlocal beam problems in order to study its performance. In general, we expect that by modeling the same nonlocal beam problems, simulation results obtained via the two approaches can achieve a good agreement. Particular attention will also be paid to studying the DO-NTBM's capability to accurately model multiscale nonlocal beams and its computational efficiency.

\begin{table}[ht!]
\caption{General parameters employed for the numerical nonlocal beam simulations. Numbers inside parentheses indicate parameters that are fixed for all the numerical simulations presented in this study. $N_x$ represents the number of mesh points in the $x$-direction, $N_y$ represents the number of mesh points in the $y$-direction, and $N_{\alpha}$ within parentheses represents the number of Gauss points when estimating DO operators (which captures the variation of nonlocality across the $y$-direction). The beam material was set to be Aluminum and its Lam\'e's parameters are also reported in the table.}
    \centering
    \begin{tabular}{||c c||c c||c c||}
    \hline
        Geometry: & $L, h$  & Mesh: & $N_x, N_y~(N_{\alpha})$ & Lam\'e's parameters: & $\mu, \lambda$ \\\cline{1-6}
        Length:  & $L~(1\mathrm{m})$ & $x$-direction: & $N_x$ & First parameter: & $\mu~(2.5967 \times 10^{10}\mathrm{Pa})$ \\\cline{1-6}
        Height:  & $h$ & $y$-direction: & $N_y~(N_{\alpha})$ & Second parameter: & $\lambda~(5.0973 \times 10^{10}\mathrm{Pa})$ \\\cline{1-6}
    \hline
    \end{tabular}
    \label{tab: 1}
\end{table}

\begin{figure}[ht!]
    \centering
    \begin{subfigure}[b]{0.49\linewidth}
        \includegraphics[width=\linewidth]{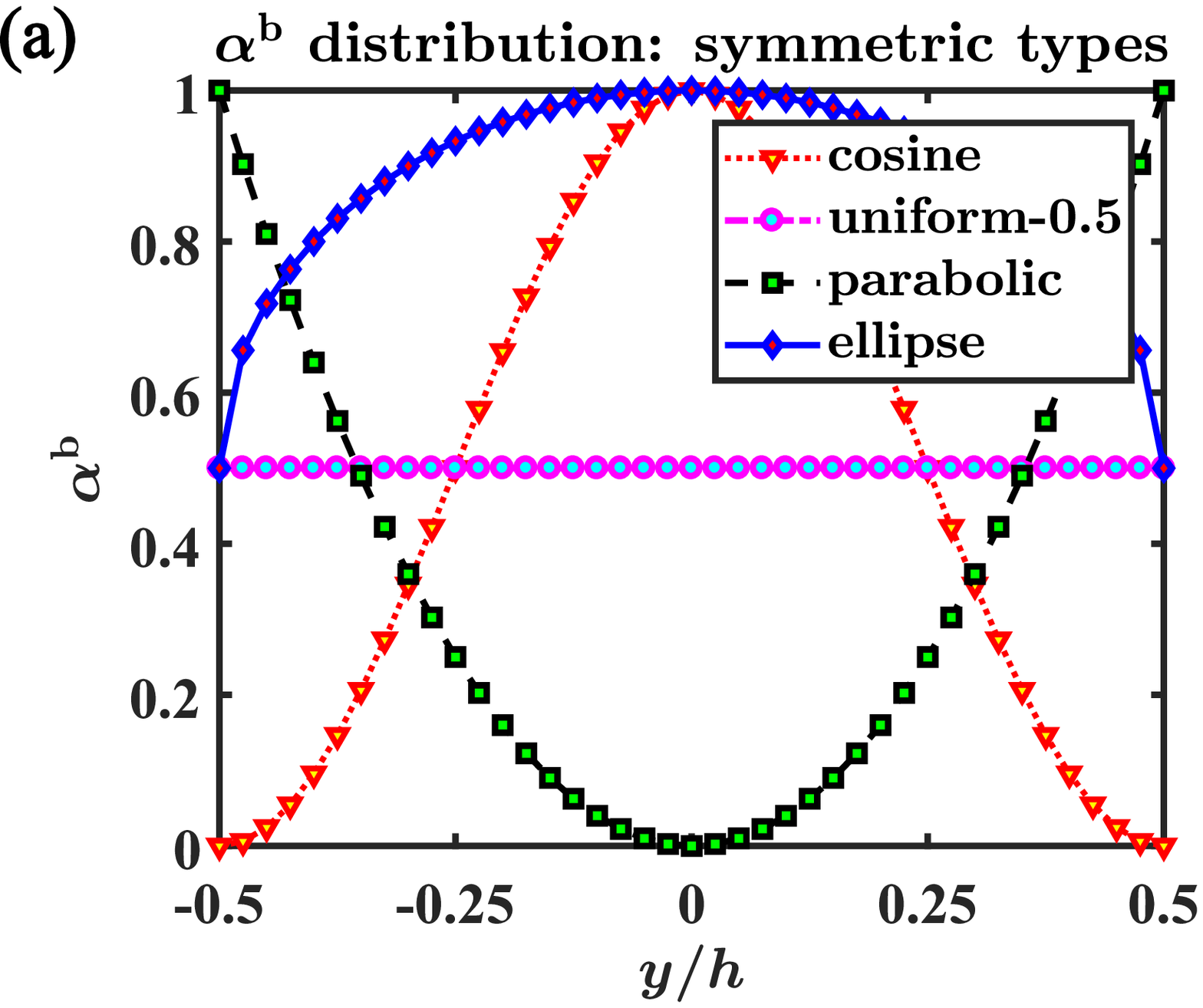}
    \end{subfigure}
    \begin{subfigure}[b]{0.49\linewidth}
        \includegraphics[width=\linewidth]{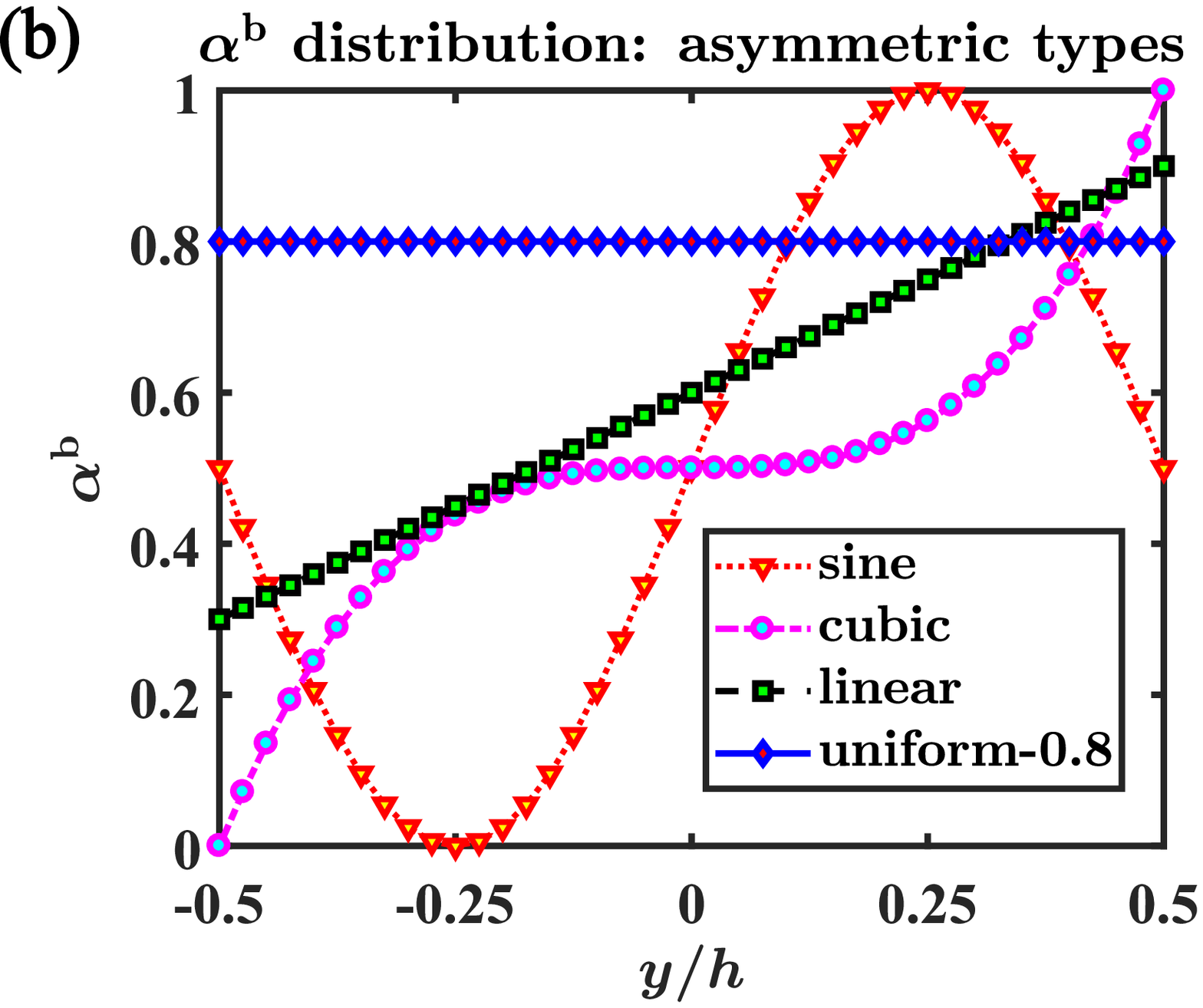}
    \end{subfigure}
    \caption{Two sets of test cases containing different distributions of nonlocal order $\alpha^{\textrm{b}}$ with respect to the normalized geometric coordinate $y$. (a) shows four distributions of nonlocal order $\alpha^{\textrm{b}}(y)$ that are symmetric about the geometric center axis $y=0$; (b) shows four distributions of $\alpha^{\textrm{b}}(y)$ that are asymmetric about the geometric center axis $y=0$. The schematic of a generic heterogeneous order distribution across the beam thickness is also included in Fig.~\ref{fig: Overview}. Detailed information of these distributions are provided in SM~\S3. Note that a uniform distribution is considered in each type of distributions ($\alpha^{\textrm{b}}(y)=0.5$ in (a) and $\alpha^{\textrm{b}}(y)=0.8$ in (b)) for reference.}
    \label{fig: alpha}
\end{figure}

Before presenting detailed simulation results, we introduce the basic configurations used to investigate the behavior of nonlocal beams. Table~(\ref{tab: 1}) lists the general parameters used in the numerical simulations. Note that both the height ($h$) and mesh ($N_x$ and $N_y~(N_\alpha)$) parameters are not predetermined so that studies involving different geometry and discretization can be conducted. For what concerns the configurations of distributed nonlocality, we primarily consider two different types of distributions (see Fig.~\ref{fig: alpha}), namely symmetric (Fig.~\ref{fig: alpha}(a)) and asymmetric (Fig.~\ref{fig: alpha}(b)) types. Following the introduction of these two types of order distributions, four different beam problems can be considered. The four problems are detailed here below:
\begin{equation}\label{eq: beam problems}
\begin{aligned}
    \bm{\mathrm{Problem~1}}\bm{:}~ &L \times h:~ 1\mathrm{m} \times 0.05\mathrm{m};~ N_x \times N_y~(N_{\alpha}):~ 901 \times 46~(46);~ q:~ 2 \times 10^7\mathrm{N}\\
    \bm{\mathrm{Problem~2}}\bm{:}~ &L \times h:~ 1\mathrm{m} \times 0.10\mathrm{m};~ N_x \times N_y~(N_{\alpha}):~ 451 \times 46~(46);~ q:~ 1 \times 10^8\mathrm{N}\\
    \bm{\mathrm{Problem~3}}\bm{:}~ &L \times h:~ 1\mathrm{m} \times 0.15\mathrm{m};~ N_x \times N_y~(N_{\alpha}):~ 301 \times 46~(46);~ q:~ 5 \times 10^8\mathrm{N}\\
    \bm{\mathrm{Problem~4}}\bm{:}~ &L \times h:~ 1\mathrm{m} \times 0.20\mathrm{m};~ N_x \times N_y~(N_{\alpha}):~ 226 \times 46~(46);~ q:~ 1 \times 10^9\mathrm{N}\\
\end{aligned}
\end{equation}
The above problems differentiate from each other based on beam geometry, discretization, and external loading. By applying various types of order distributions as well as beam configurations, we aim at testing the modeling approach on different mechanical problems. Numerical algorithms used to perform the simulations for both DO-ANET and DO-NTBM are presented in SM~\S2.

\subsection{Elastostatic analysis of multiscale nonlocal beams}\label{sssec: elastostatic analysis}
Based on the above configurations, numerical simulations are performed. Figs.~(\ref{fig: u_DO-ANET}-\ref{fig: sigma_U_xy}) show results for the different nonlocal beams including displacement and stress distributions obtained via both DO-ANET and DO-NTBM theories. Specifically, Fig.~\ref{fig: u_DO-ANET} shows the full 2D distribution of the displacement field predicted by DO-ANET. By taking the DO-ANET's results as reference, the transverse displacement of the nonlocal beams predicted by the DO-NTBM (Fig.~\ref{fig: uy}) can be evaluated. The distribution of both shear stress $\sigma_{xy}$ and shear energy $U_{xy}$ predicted by both approaches are shown in Figs.~(\ref{fig: sigma_sine},\ref{fig: sigma_U_xy}); these results provide further insight in the ability of the method to capture multiscale characteristics. Table~(\ref{tab: 2}) lists specific values of the Timoshenko beam parameters ($h_c$ and $\chi$) obtained from simulations.

At a first glance, these results suggest that: 1) the transverse displacement field ($u_y$) predicted by both the DO-ANET and the DO-NTBM are generally in good agreement, and 2) the overall distribution of the stress fields recovered from the reduced 1D model (DO-NTBM) coincide with their counterparts obtained by the 2D model (DO-ANET). Further analyses of these results lead to the following important observations and remarks (note that the first remark in the following is presented to show DO-ANET's effectiveness and based on that, the second and third remarks are presented together to further demonstrate the DO-NTBM's ability to model multiscale nonlocal effects):

\begin{figure}[ht!]
    \centering
    \begin{subfigure}[b]{0.49\linewidth}
        \includegraphics[width=\linewidth]{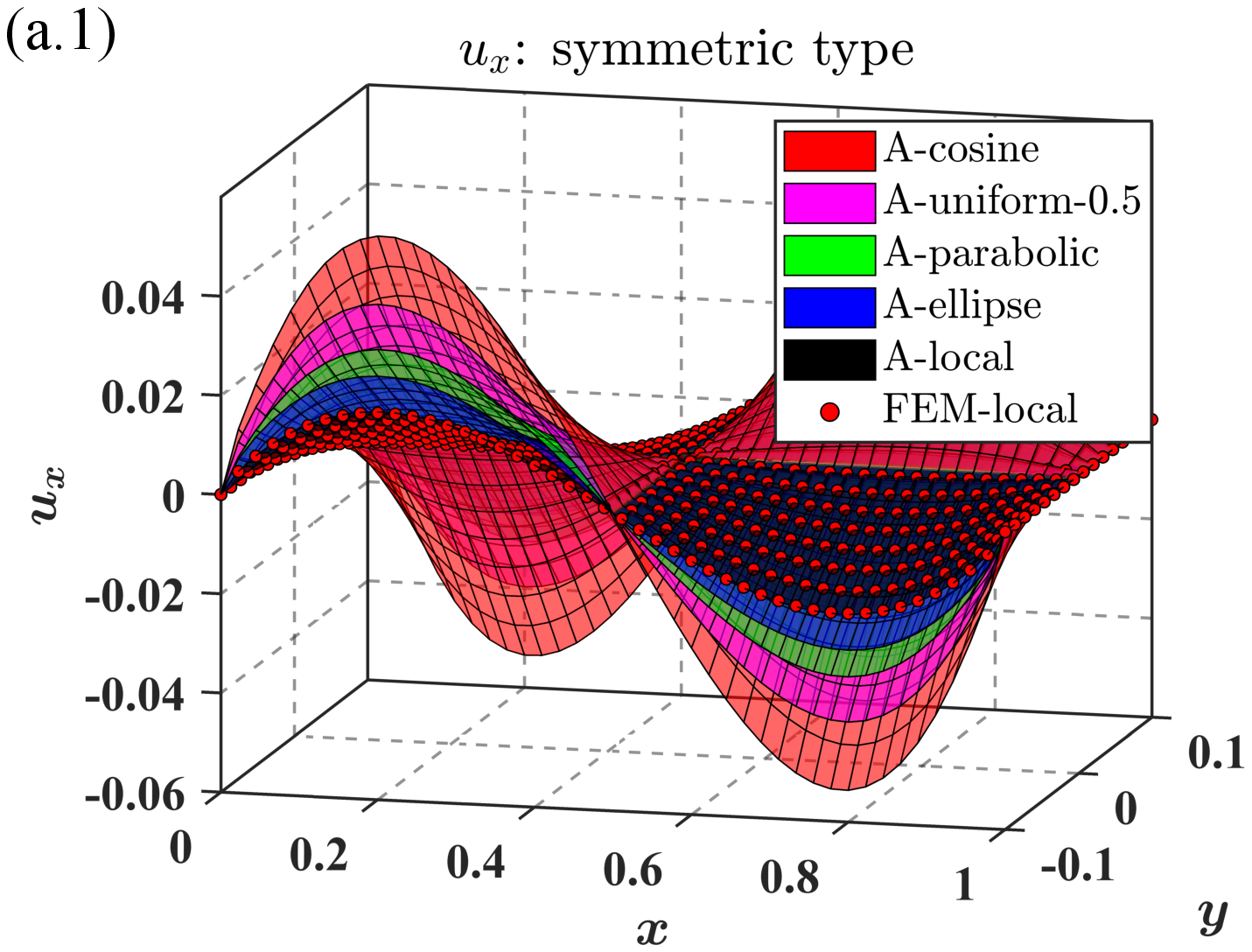}
    \end{subfigure}
    \begin{subfigure}[b]{0.49\linewidth}
        \includegraphics[width=\linewidth]{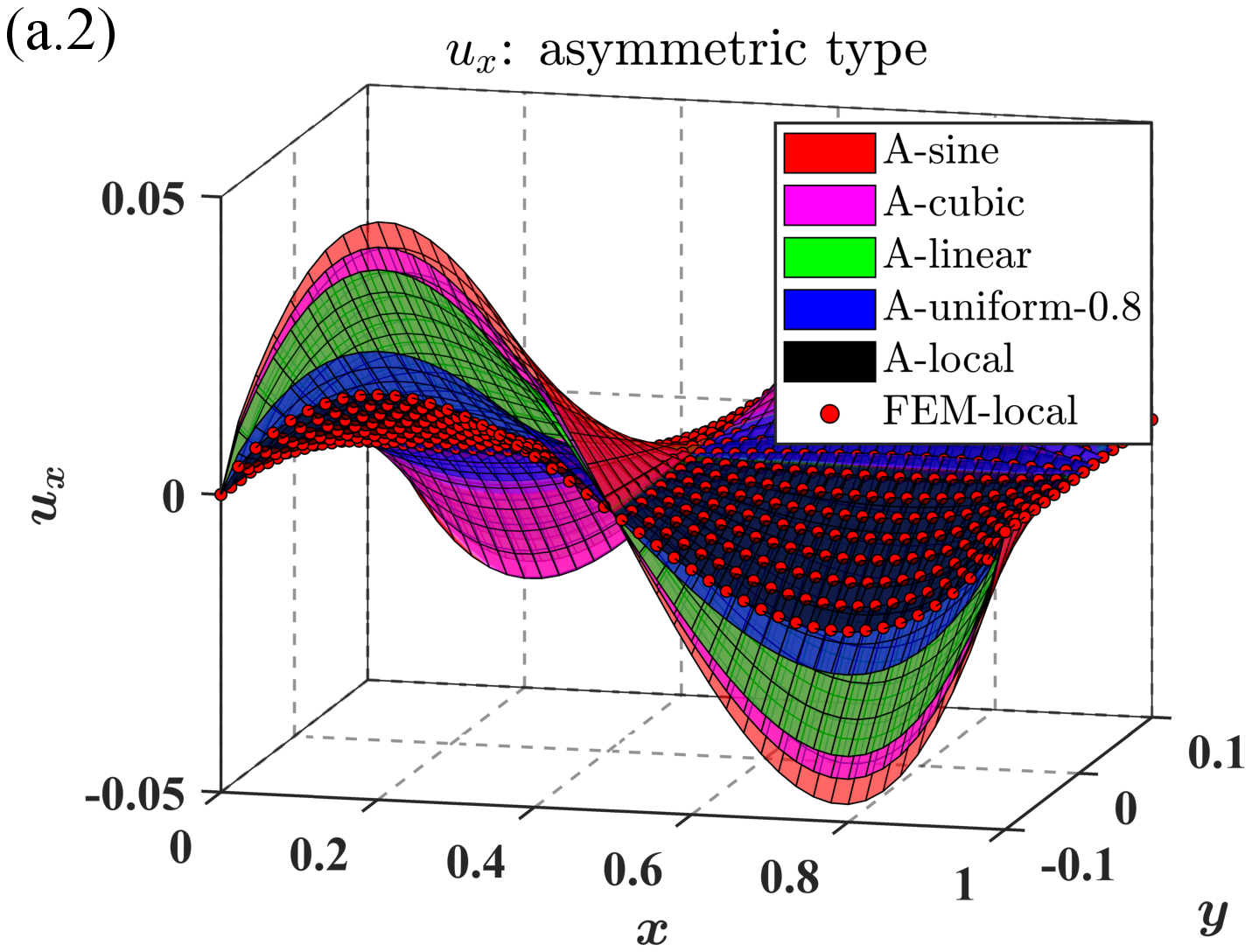}
    \end{subfigure}
    \begin{subfigure}[b]{0.49\linewidth}
        \includegraphics[width=\linewidth]{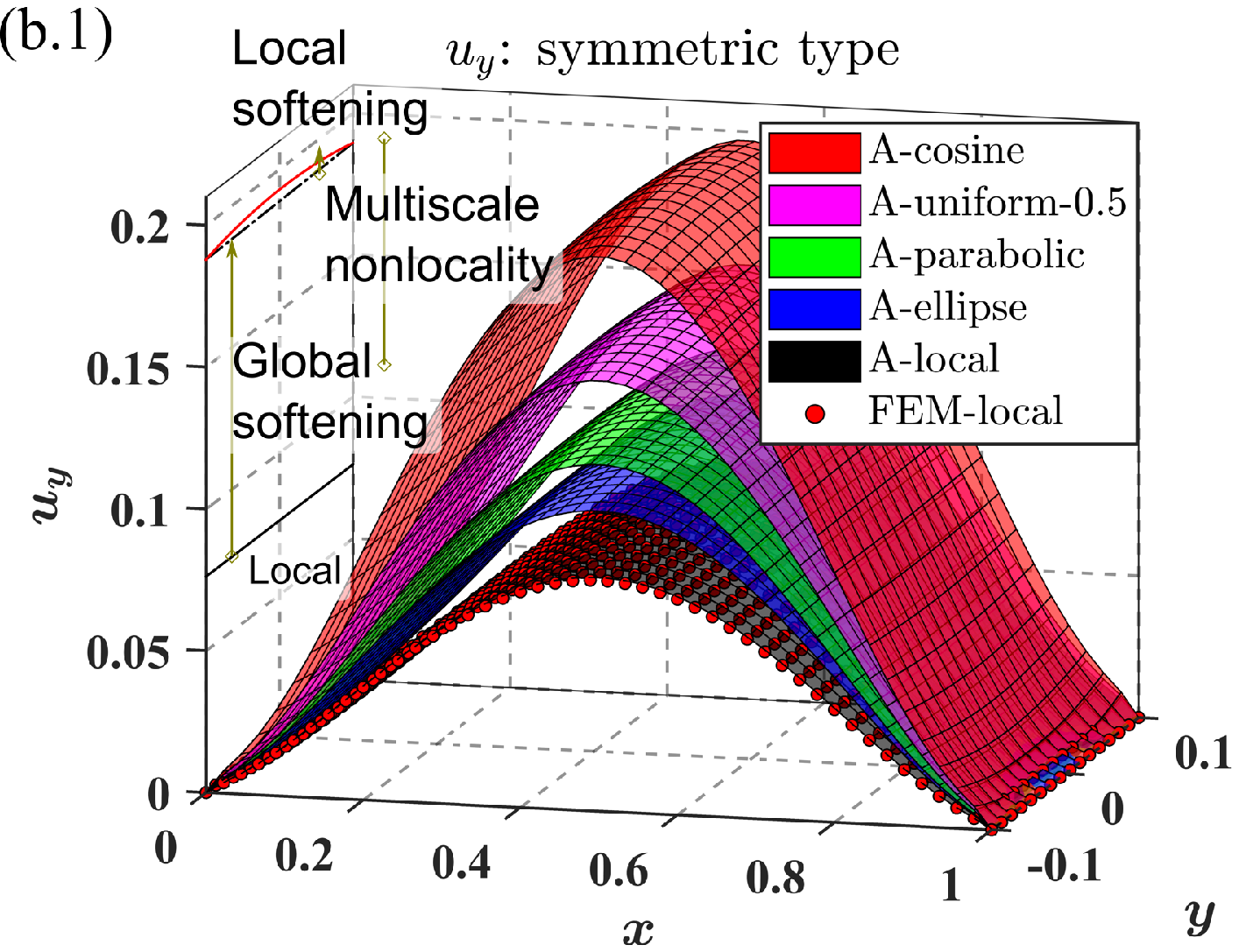}
    \end{subfigure}
    \begin{subfigure}[b]{0.49\linewidth}
        \includegraphics[width=\linewidth]{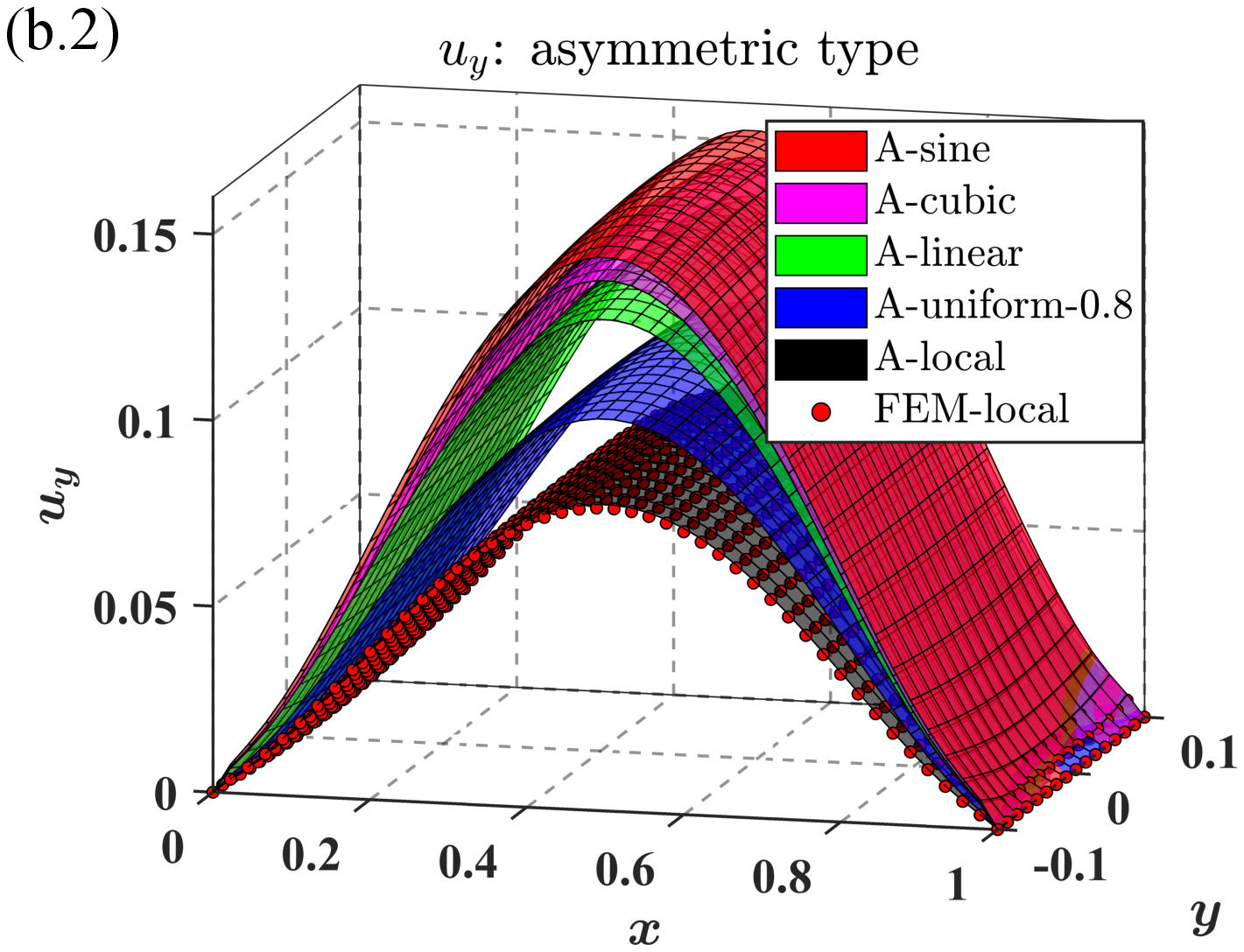}
    \end{subfigure}
    \caption{Full 2D distribution of the displacement field components $u_x$ and $u_y$: (a.1) and (a.2) show the $x$ displacement $u_x$ with symmetric and asymmetric type order distributions; (b.1) and (b.2) show the displacement $u_y$ with symmetric and asymmetric type order distributions. Black surface plots and red dotted scatter plots represent the results of simulations for local systems (where $\alpha^{\textrm{b}}(y)=1$) computed by the 2D DO-ANET (labeled `A') and the finite element method via COMSOL Multiphysics (labeled `FEM-local'), respectively. $\bm{\mathrm{Problem~4}}$ is used for this set of simulations. Note that the number of mesh points is reduced to improve data visualization.}
    \label{fig: u_DO-ANET}
\end{figure}

\begin{enumerate}[leftmargin=*]
    \item[$\bm{(1)}$] The DO-ANET can capture the detailed 2D field distributions of the nonlocal beams. Fig.~\ref{fig: u_DO-ANET} shows the two components of the displacement field, $u_x$ and $u_y$, as predicted by the 2D DO-ANET. We observe that, the displacement responses in both the $x$- and $y$- directions increase to varying degrees when considering different order distribution $\alpha^{\textrm{b}}(y)$. We highlight that the increase of displacement is consistent with the fundamental material softening characteristic of fractional nonlocal elasticity and, more in general, of nonlocal mechanics. Further observations reinforce the understanding of the relationships between the material softening effect and the distribution of nonlocal effects represented by $\alpha^{\textrm{b}}(y)$: as $\alpha^{\textrm{b}}(y)$ decreases in magnitude, the nonlocal effect becomes more pronounced and the material exhibits an increasing softening behavior. Consider, for example, two symmetric distributions $\bm{\mathrm{'uniform\text{-}0.5'}}$ and $\bm{\mathrm{'ellipse'}}$ shown in Fig.~\ref{fig: alpha}(a). The order of nonlocality $\alpha^{\textrm{b}}(y)$ in the $\bm{\mathrm{'uniform\text{-}0.5'}}$ distribution is always smaller than the order in the $\bm{\mathrm{'ellipse'}}$ distribution $\forall y \in [-h/2,h/2]$; this indicates that nonlocal beams following the $\bm{\mathrm{'uniform\text{-}0.5'}}$ distribution should always be softer, and eventually lead to larger displacements, under the same external load. Simulation results in Fig.~\ref{fig: u_DO-ANET}(a.1-b.1) show that both displacement components resulting from the $\bm{\mathrm{'uniform\text{-}0.5'}}$ distribution are always larger than their counterpart under the $\bm{\mathrm{'ellipse'}}$ distribution. This feature is clearly in line with the above theoretical interpretation and confirms that the proposed DO-ANET can indeed capture the material softening effect. Note that to further substantiate the effectiveness of this approach, simulations of the transverse bending of a fully local beam were also considered. In this limiting case, we set $\alpha^{\textrm{b}}(y)=1$ $\forall y \in [-h/2,h/2]$. The numerical results obtained by the DO-ANET (see the black surface plot) and by an independent finite element model built in the commercial package COMSOL Multiphysics (see the red dot scatter plot) are also presented and compared in Fig.~\ref{fig: u_DO-ANET}. We observe that both displacement components $u_x$ and $u_y$ computed by the two approaches are in excellent agreement, hence further substantiating the ability of DO-ANET to model the elastostatic response of 2D beams.
    
    \begin{figure}[hp]
        \centering
        \begin{subfigure}[b]{0.49\linewidth}
            \includegraphics[width=\linewidth]{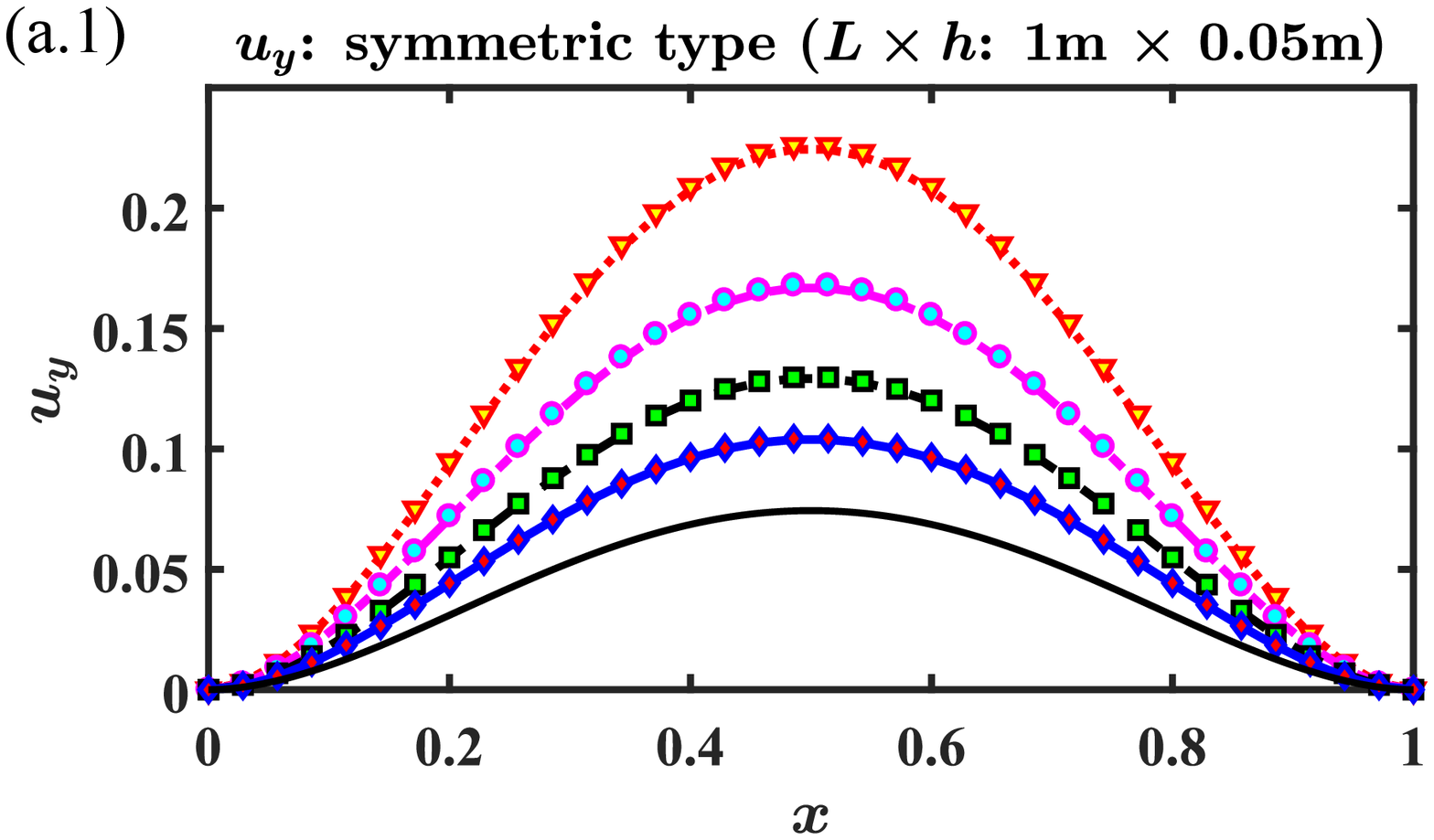}
        \end{subfigure}
        \begin{subfigure}[b]{0.49\linewidth}
            \includegraphics[width=\linewidth]{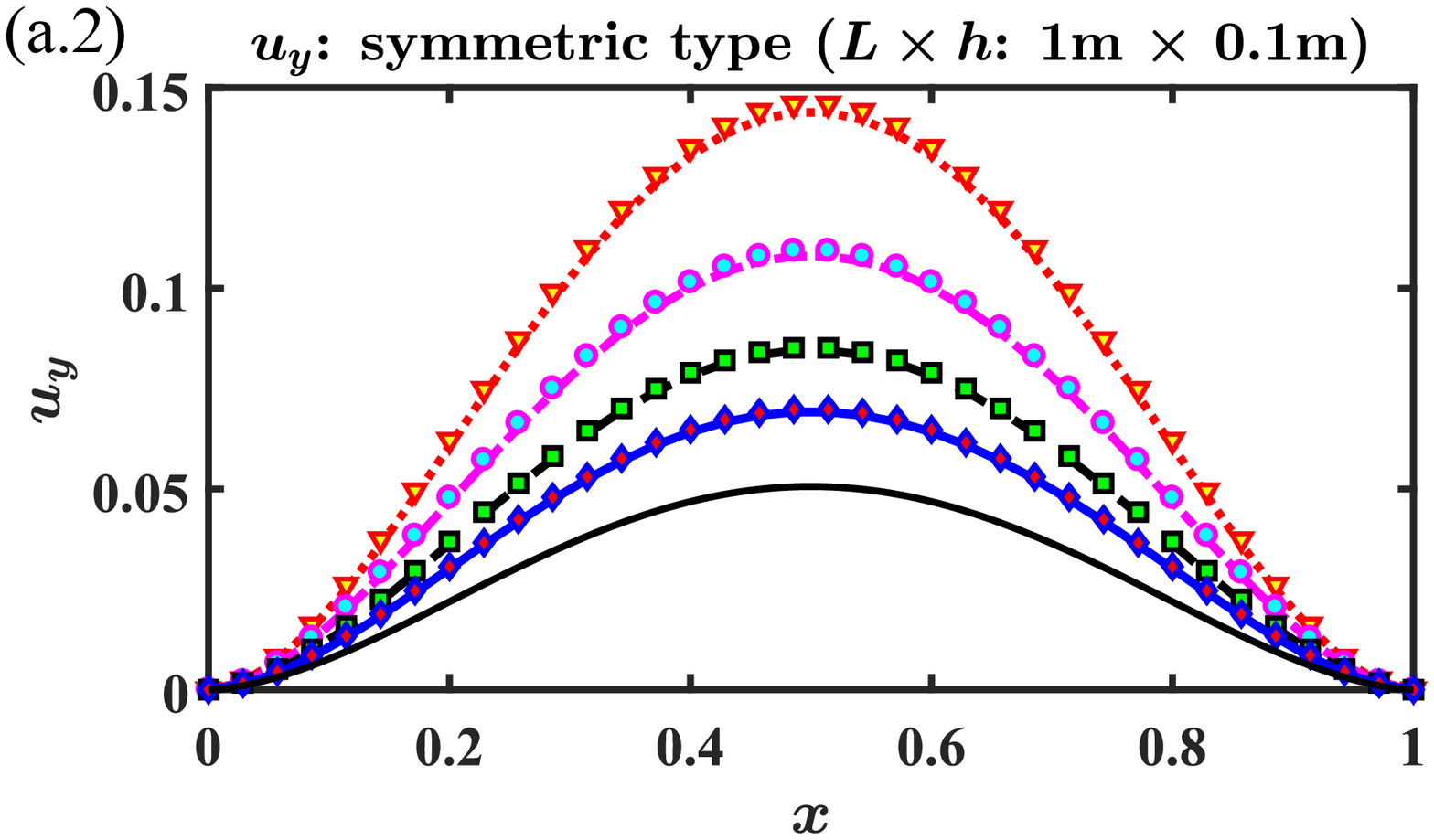}
        \end{subfigure}
        \begin{subfigure}[b]{0.49\linewidth}
            \includegraphics[width=\linewidth]{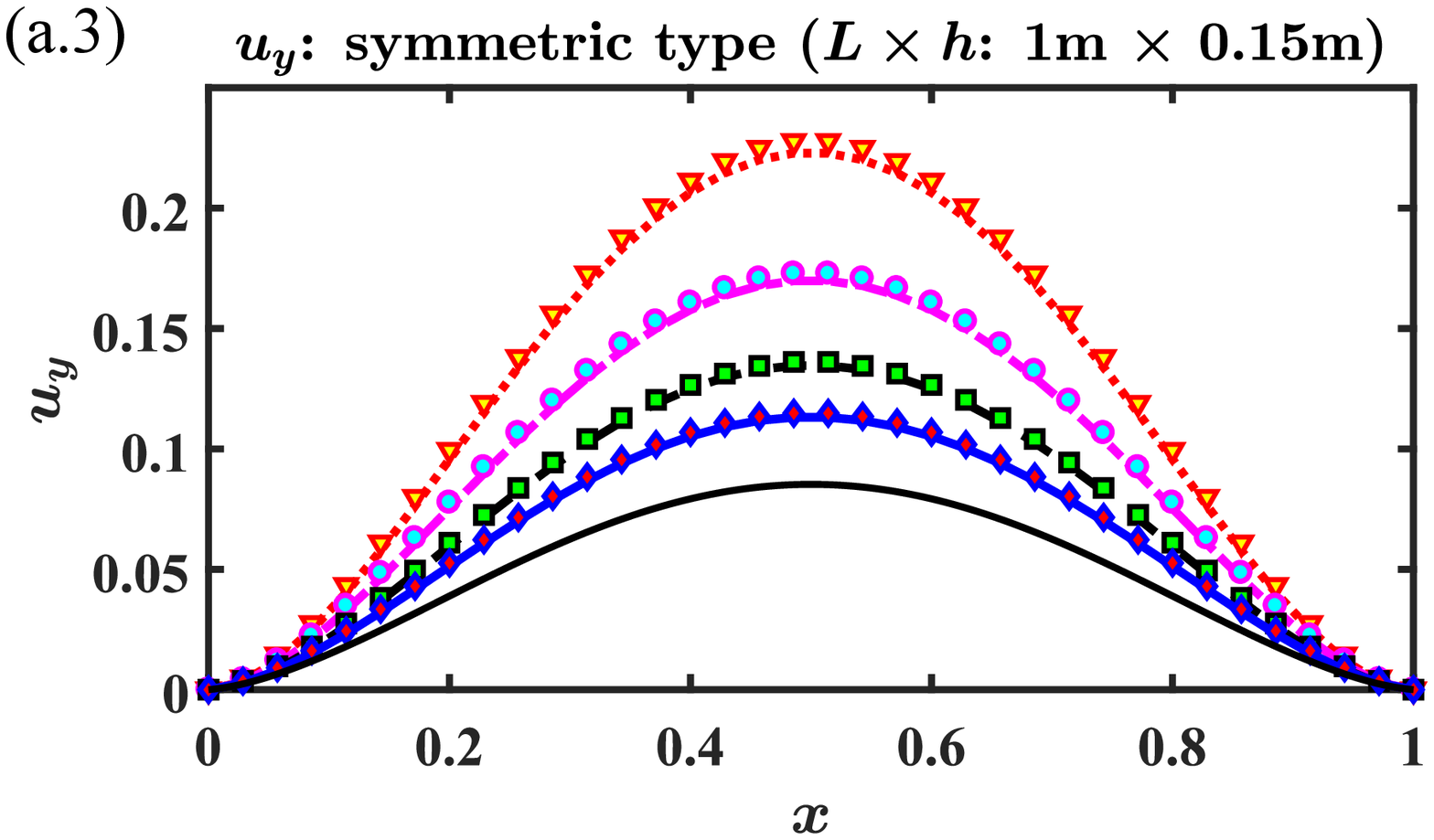}
        \end{subfigure}
        \begin{subfigure}[b]{0.49\linewidth}
            \includegraphics[width=\linewidth]{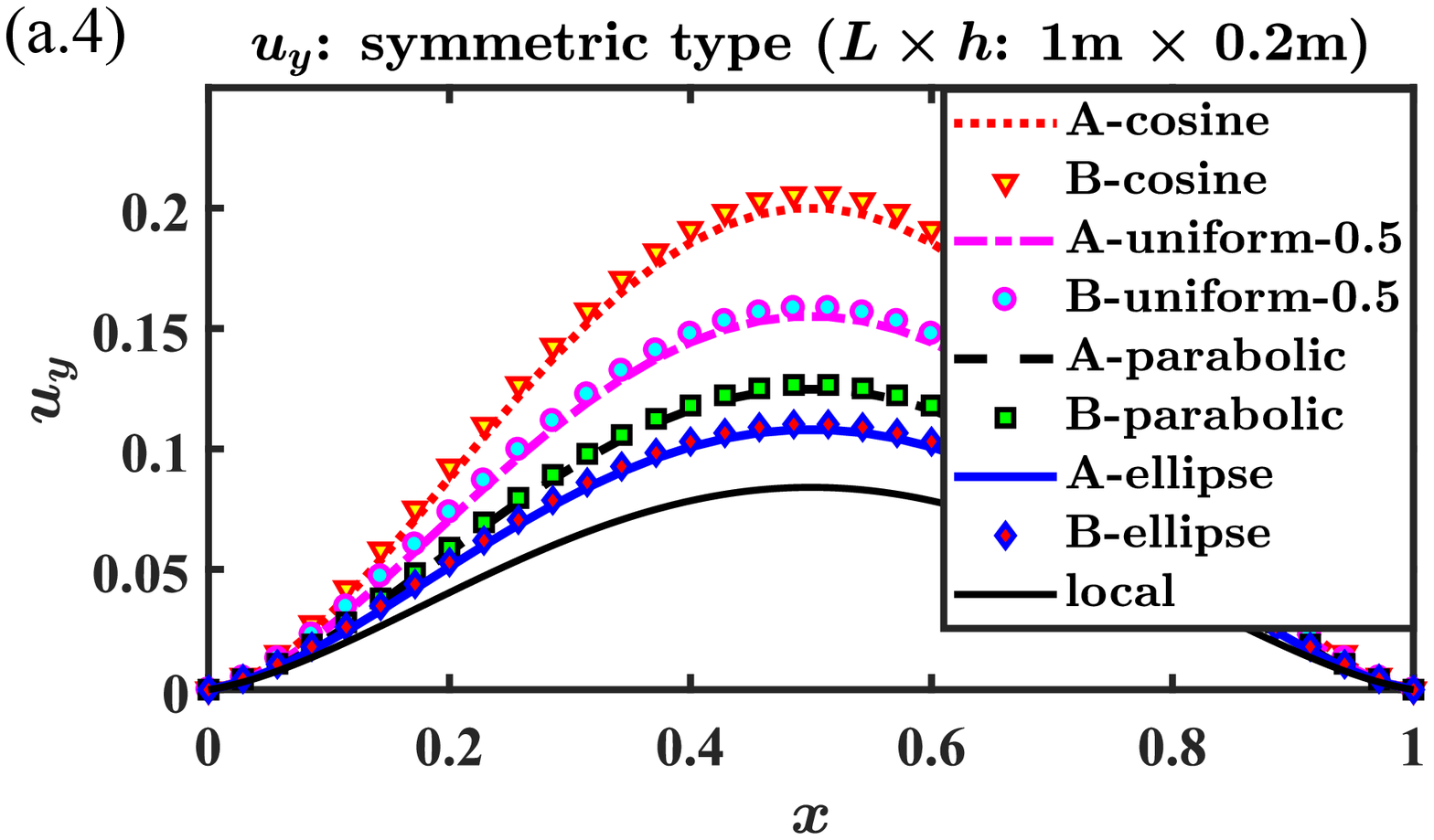}
        \end{subfigure}
        \begin{subfigure}[b]{0.49\linewidth}
            \includegraphics[width=\linewidth]{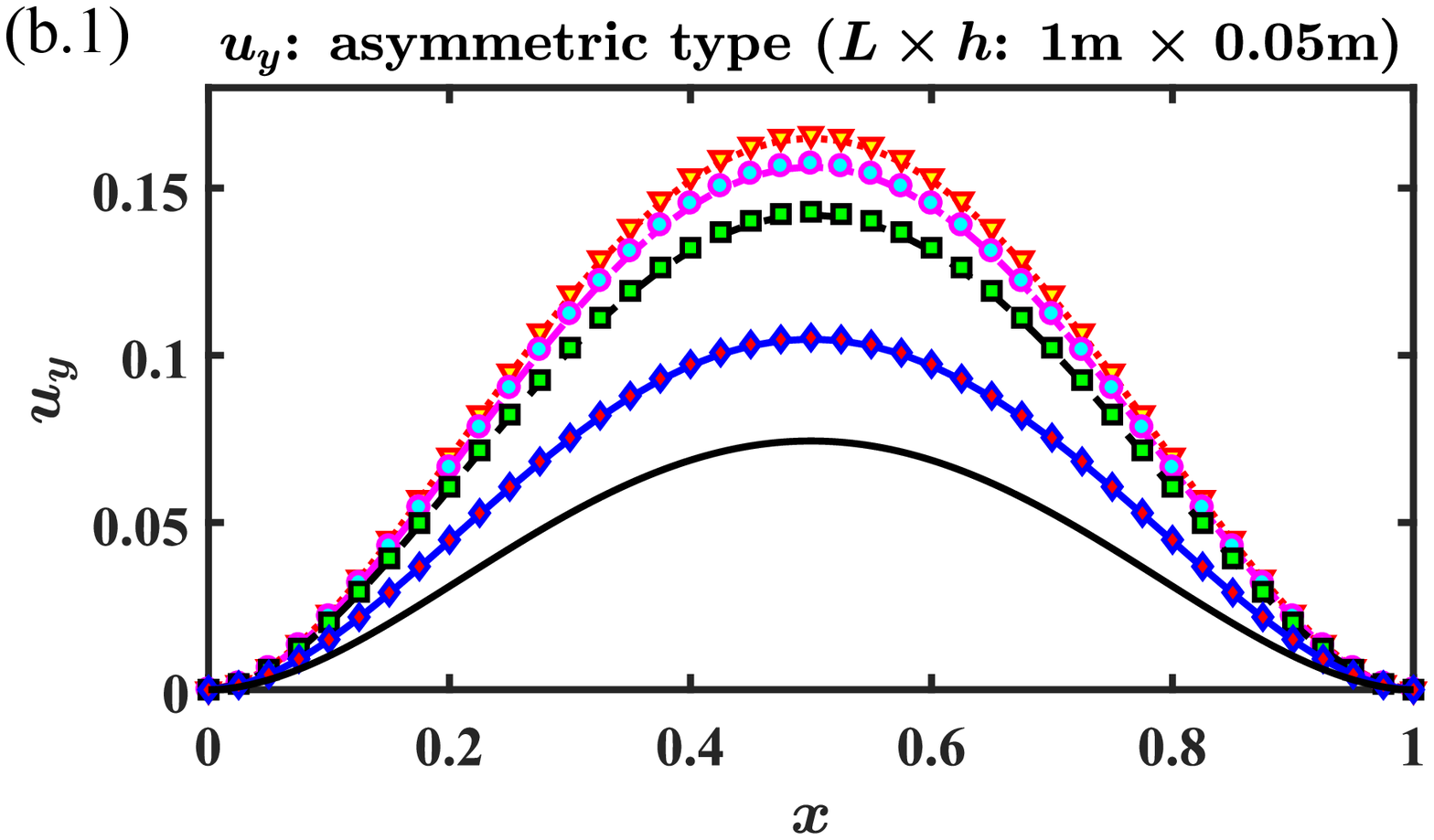}
        \end{subfigure}
        \begin{subfigure}[b]{0.49\linewidth}
            \includegraphics[width=\linewidth]{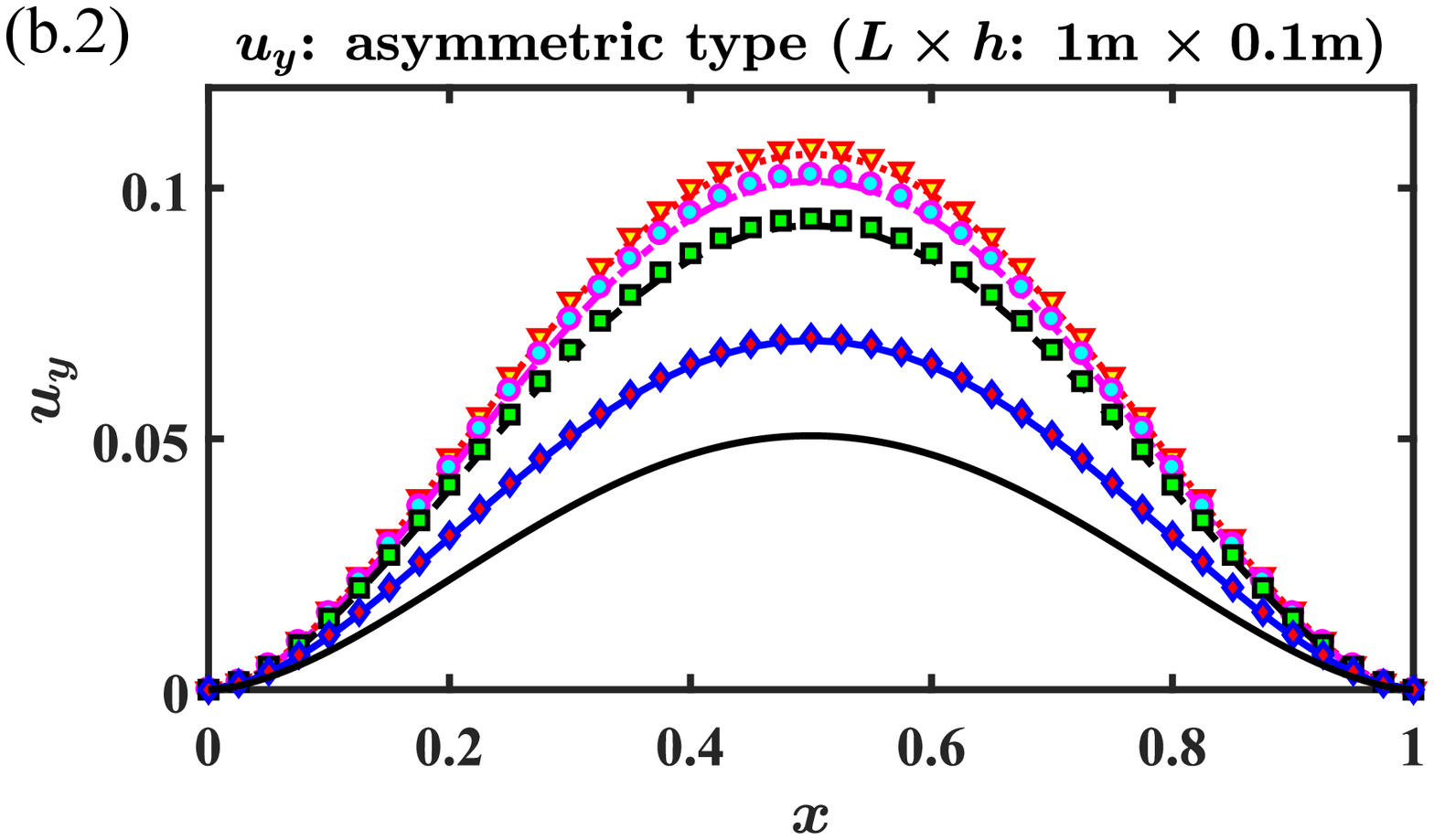}
        \end{subfigure}
        \begin{subfigure}[b]{0.49\linewidth}
            \includegraphics[width=\linewidth]{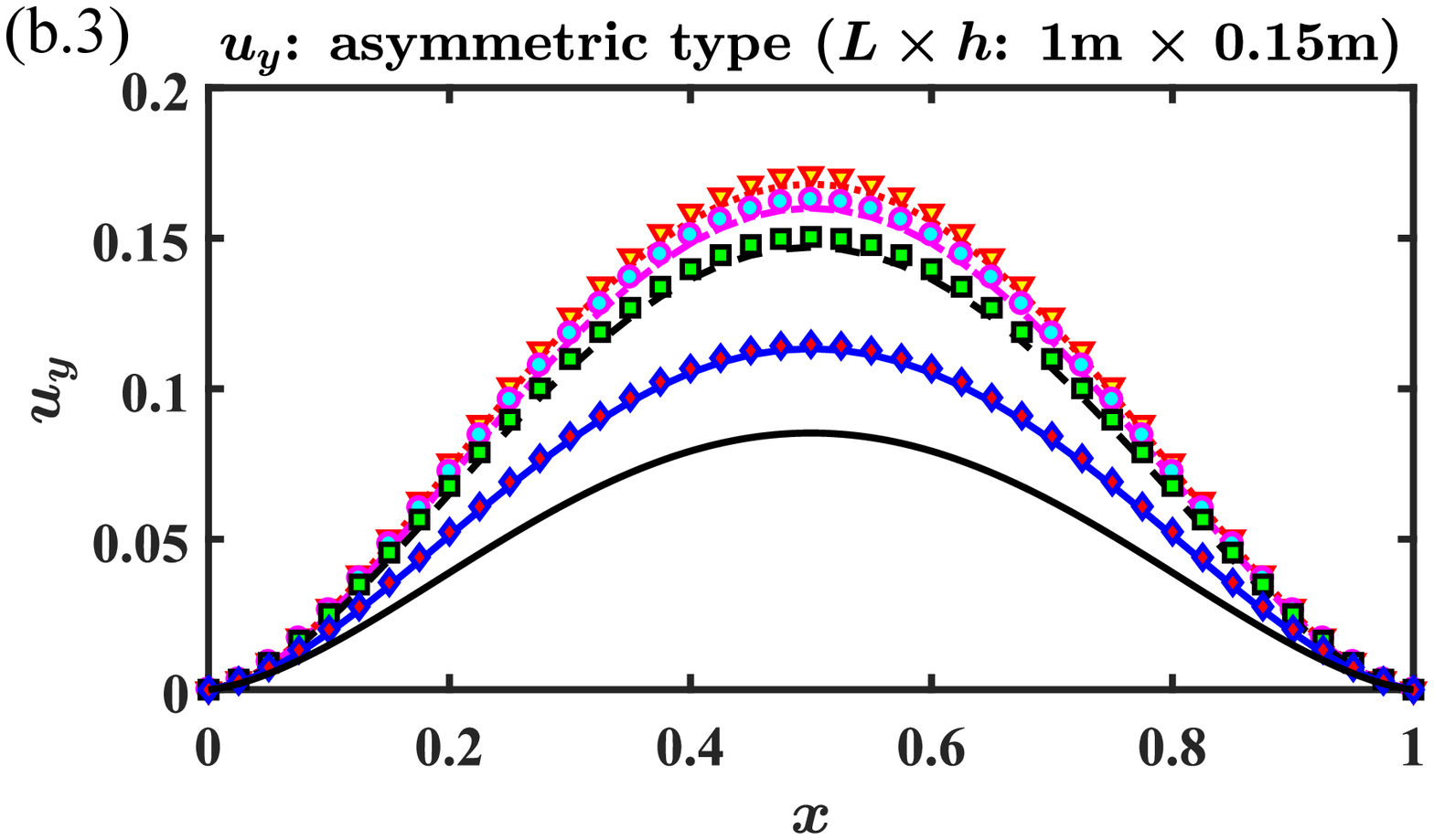}
        \end{subfigure}
        \begin{subfigure}[b]{0.49\linewidth}
            \includegraphics[width=\linewidth]{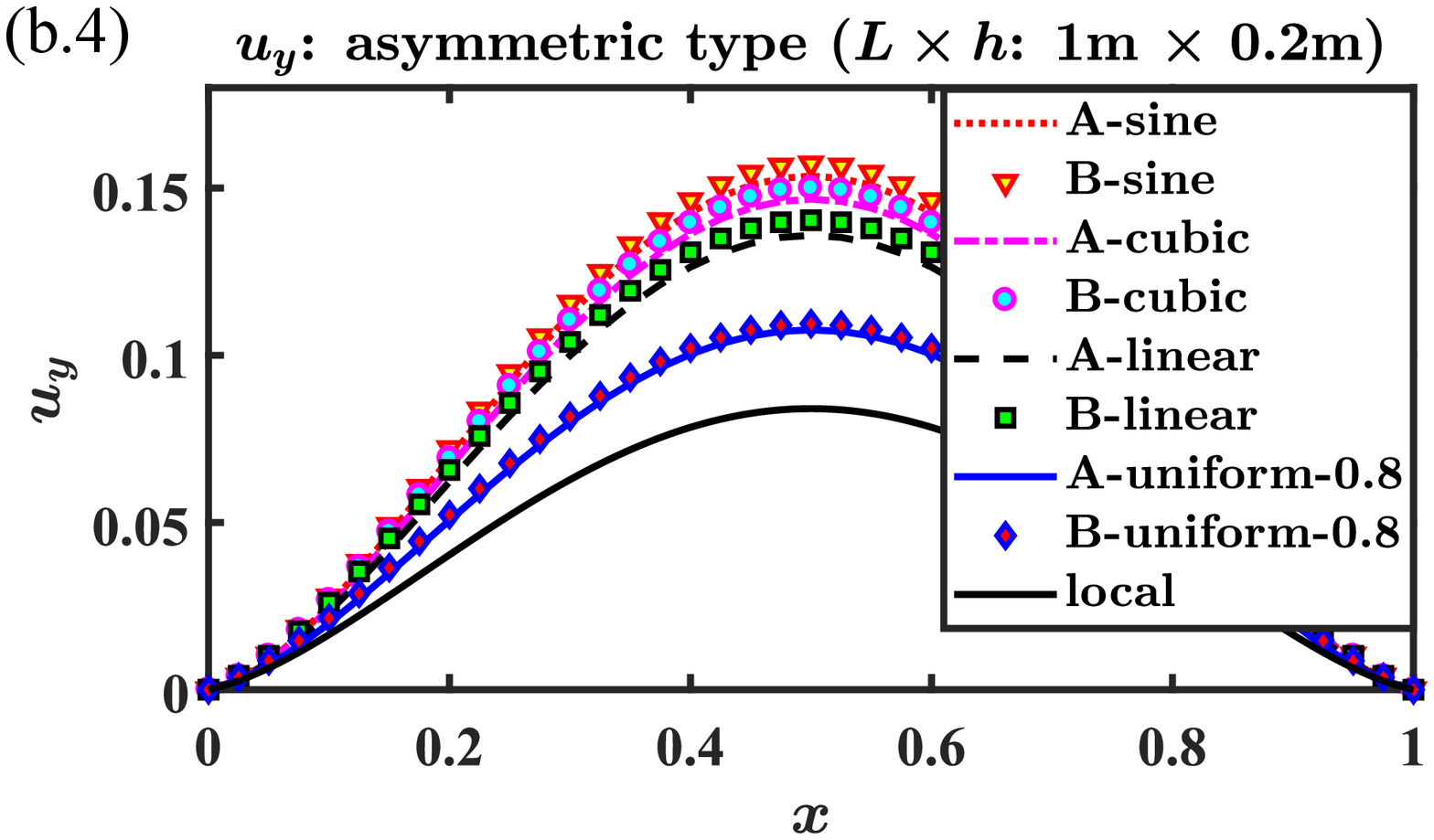}
        \end{subfigure}
        \caption{Transverse displacement of nonlocal beams under different distribution of orders. Four different beam problems listed in Eq.~(\ref{eq: beam problems}) are considered. Subfigures (a.1)-(a.4) and (b.1)-(b.4) show simulation results to the four problems using symmetric and asymmetric types of order $\alpha^{\textrm{b}}(y)$, respectively. $\bm{\mathrm{'A'}}$ and $\bm{\mathrm{'B'}}$ displayed in legends refer to simulation results obtained by DO-ANET and DO-NTBM, respectively.}
        \label{fig: uy}
    \end{figure}
    
    \begin{table}[ht!]
    \centering
    \begin{tabular}{c | c | c | c | c | c | c | c | c | c}
    \hline
    \hline
        \multirow{2}{*}{} & \multirow{2}{*}{$h$(m)} & \multicolumn{4}{c|}{\textbf{symmetric distribution}} & \multicolumn{4}{c}{\textbf{asymmetric distribution}} \\\cline{3-10}
        & & $\bm{\mathrm{uniform\text{-}0.5}}$ & $\bm{\mathrm{parabolic}}$ & $\bm{\mathrm{cosine}}$ & $\bm{\mathrm{ellipse}}$ & $\bm{\mathrm{sine}}$ & $\bm{\mathrm{cubic}}$ & $\bm{\mathrm{linear}}$ & $\bm{\mathrm{uniform\text{-}0.8}}$\\ \hline
        \hline
        \multirow{4}{*}{$h_c$(m)} & 0.05 & 0.0000 & 0.0000 & 0.0000 & 0.0000 & 0.0053 & 0.0037 & 0.0038 & 0.0000 \\\cline{2-10}
        & 0.1 & 0.0000 & 0.0000 & 0.0000 & 0.0000 & 0.0107 & 0.0074 & 0.0075 & 0.0000 \\\cline{2-10}
        & 0.15 & 0.0000 & 0.0000 & 0.0000 & 0.0000 & 0.0160 & 0.0111 & 0.0113 & 0.0000 \\\cline{2-10}
        & 0.2 & 0.0000 & 0.0000 & 0.0000 & 0.0000 & 0.0213 & 0.0148 & 0.0150 & 0.0000 \\
        \hline
        \hline
        \multirow{4}{*}{$\chi$} & 0.05 & 0.8344 & 0.8982 & 0.7831 & 0.7828 & 0.8393 & 0.8561 & 0.8098 & 0.8343 \\\cline{2-10}
        & 0.1 & 0.8338 & 0.8984 & 0.7821 & 0.7850 & 0.8320 & 0.8478 & 0.8001 & 0.8343 \\\cline{2-10}
        & 0.15 & 0.8338 & 0.8984 & 0.7812 & 0.7862 & 0.8269 & 0.8416 & 0.7951 & 0.8343 \\\cline{2-10}
        & 0.2 & 0.8339 & 0.8983 & 0.7805 & 0.7871 & 0.8234 & 0.8373 & 0.7904 & 0.8343 \\\cline{2-10}
    \hline
    \hline
    \end{tabular}
    \caption{Values of parameters obtained from DO Timoshenko beam simulations in Fig.~\ref{fig: uy}. $h_c$ is strongly affected by the distribution type (symmetric/asymmetric) of the nonlocal order. $\chi$ also changes with the different distributions but does not vary significantly. Both parameters are not sensitive to the beam geometry.}
    \label{tab: 2}
    \end{table}
    
    \item[$\bm{(2)}$] DO-NTBM can well capture the overall effect of the heterogeneous distribution of nonlocal order $\alpha^{\textrm{b}}(y)$ at the macro scales. Unlike the DO-ANET which is a 2D modeling approach, the DO-NTBM is a 1D model in nature (see the governing equations Eq.~(\ref{eq: Timoshenko equation-2})). Although in a beam formulation the transverse direction is typically reduced (due to kinematic assumptions), our simulations highlight that by leveraging the DO formulation the through-the-thickness nonlocal effects (captured by $\alpha^{\textrm{b}}(y)$) can still be accounted for. Similar to the DO-ANET, the same material softening effect can be predicted by the DO-NTBM. Specifically, simulation results in Fig.~\ref{fig: uy}(a.1-a.4) have shown that for all the four different beam problems, the transverse displacement in the $\bm{\mathrm{'uniform\text{-}0.5'}}$ distribution is always larger than its counterpart given by the $\bm{\mathrm{'ellipse'}}$ distribution. With the overall order-sensitive softening effects being predicted, further observation suggests that the difference of $u_y$ at the macro scale can be also quantitatively predicted by the DO-NTBM. Simulation results obtained by both the DO-ANET and the DO-NTBM are generally in good agreement for different order distributions (see Fig.~\ref{fig: alpha}) and beam configurations (see Eq.~(\ref{eq: beam problems})). The relative difference between $u_y$ caused by $\alpha^{\textrm{b}}(y)$ is accurately predicted by both approaches. Although a slight inconsistency of $u_y$ is found when $h=0.2$m with asymmetric types of order distribution (see Fig.~\ref{fig: uy}(b.4)), we note that the error in this case not only originates from $\alpha^{\textrm{b}}(y)$ but also from the beam geometry with an aspect ratio $L/h=5$ that exceeds the range of applicability of Timoshenko beams. Factoring in these latter aspects, the agreement between the results produced by the DO-ANET and the DO-NTBM holds in a general sense, hence substantiating the fact that 1) the DO formulation can accurately capture the variation of mechanical responses due to the heterogeneously distributed nonlocal order, and 2) the Timoshenko formulation is still suitable to describe the overall nonlocal elastic behavior of thick beams.
    
    \begin{figure}[ht!]
        \centering
        \includegraphics[width=\linewidth]{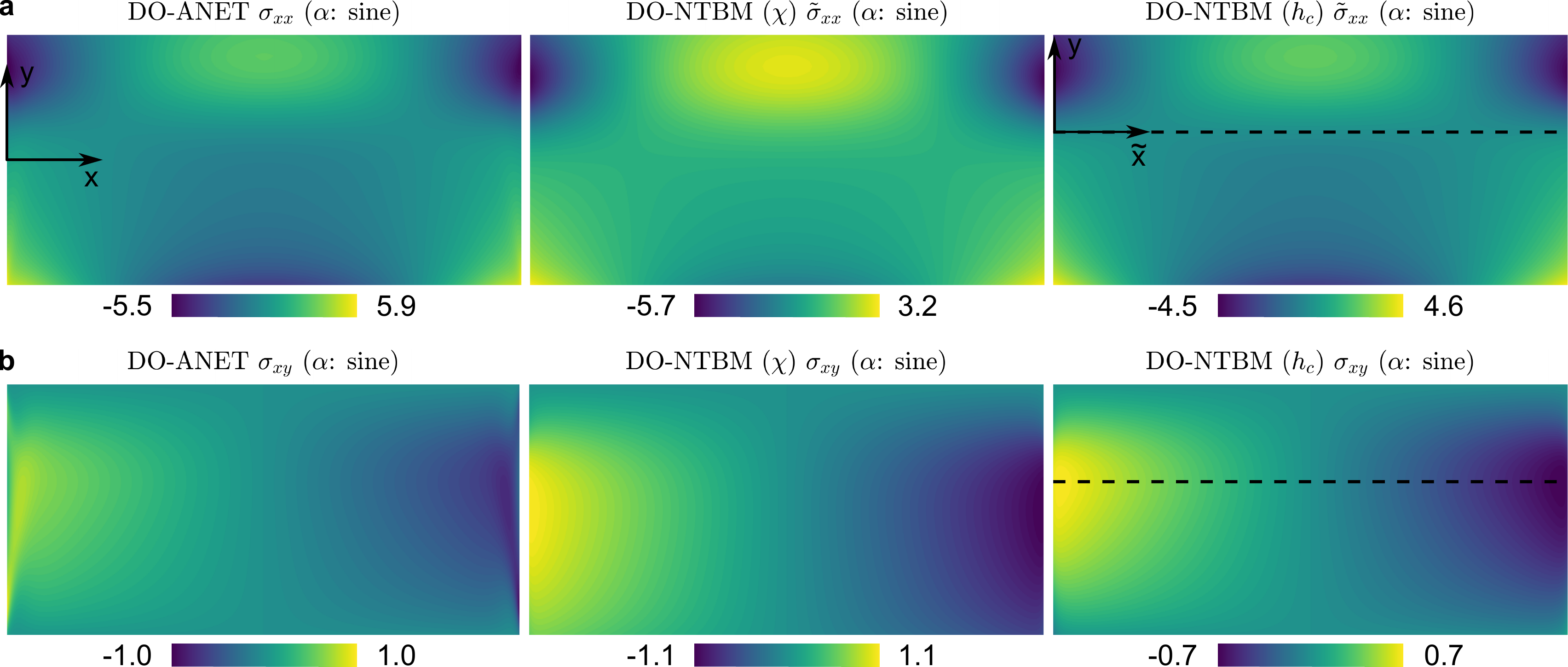}
        \caption{Fully resolved 2D distribution of stress (in GPa) within the nonlocal beam. (a) shows simulation results of the normal stress $\sigma_{xx}$ and (b) shows the shear stress $\sigma_{xy}$. For both stress components, the following three simulation approaches are used: 1) DO-ANET, 2) DO-NTBM with $\chi$ only, and 3) DO-NTBM with $h_c$ only. The black dashed lines shown in the results predicted by DO-NTBM with $h_c$ only indicate the position of the physical center axis. $\bm{\mathrm{Problem~2}}$ and the $\bm{\mathrm{'sine'}}$ distribution of $\alpha^{\textrm{b}}(y)$ are used for simulation.}
        \label{fig: sigma_sine}
    \end{figure}
    
    \item[$\bm{(3)}$] Implementation of the two extra parameters $h_c$ and $\chi$ and of the associated auxiliary equations enables the DO-NTBM the ability to capture the effect of the heterogeneous nonlocal order $\alpha^{\textrm{b}}(y)$ at the micro scales. Detailed analysis of results in Figs.~\ref{fig: uy}-\ref{fig: sigma_U_xy} and Table~(\ref{tab: 2}) leads to the following observations and remarks:
    \begin{itemize}[leftmargin=*]
        \item $h_c$ is strongly affected by the order distribution. Specifically, as it shows in Table~(\ref{tab: 2}), $h_c$ are all zero for simulations with symmetric $\alpha^{\textrm{b}}(y)$ and all nonzero for asymmetric $\alpha^{\textrm{b}}(y)$ (except $\bm{\mathrm{'uniform\text{-}0.8'}}$). The strong correlation between $h_c$ and the symmetry of $\alpha^{\textrm{b}}(y)$ leads to the following interpretation.
        According to Eqs.~(\ref{eq: displacement-Timoshenko}-\ref{eq: sigma-Timoshenko}), $h_c$ mainly captures the microscale asymmetric behavior of physical quantities (such as displacements, strains, and stresses) imposed by the heterogeneous distribution of $\alpha^{\textrm{b}}(y)$. For symmetric types of distribution, $\alpha^{\textrm{b}}(y)$ is symmetrically distributed about the geometric center axis $y=0$ such that the variation of physical properties brought by the heterogeneous distributed-order at the top half section ($y>0$) and the bottom half section ($y<0$) are the same. In this regard, the physical center axis coincides with the geometric center axis ($h_c=0$, see Table~(\ref{tab: 2})). For asymmetric types of distribution, since the variation of physical properties at the upper and bottom sections is not the same, the physical center axis deviates from the geometric center axis ($h_c \neq 0$, see Table~(\ref{tab: 2})). Straightforward results supporting the interpretation can be found in Fig.~\ref{fig: sigma_sine} where the transverse distribution of normal and shear stress obtained by the DO-ANET, the DO-NTBM with $h_c$, and the DO-NTBM without $h_c$ are presented. Specifically, we observe that for $\bm{\mathrm{'sine'}}$ distribution of $\alpha^{\textrm{b}}(y)$, both normal and shear stress fields in DO-ANET simulations are not symmetric and clearly deviate from the geometric center axis $y=0$ (see the left two subfigures in Fig.~\ref{fig: sigma_sine}). While the same feature can be reproduced in DO-NTBM simulations with $h_c$ (the position of physical center axes are marked explicitly by black dashed lines in the right two subfigures in Fig.~\ref{fig: sigma_sine}), DO-NTBM simulations do not involve $h_c$ (see the middle two subfigures in Fig.~\ref{fig: sigma_sine}) and fail to capture the deviation.
        
        Additional analysis reveals a positive correlation between $h_c$ and $\alpha^{\textrm{b}}(y)$. Particularly, the more asymmetric $\alpha^{\textrm{b}}(y)$ is, the farther the physical center axis will deviate from $y=0$. Specific simulation results for the $\bm{\mathrm{'sine'}}$ and $\bm{\mathrm{'linear'}}$ distribution cases further justify this argument. As it is shown in Fig.~\ref{fig: alpha}(b), the $\bm{\mathrm{'sine'}}$ curve ranges from $\alpha=0$ to $\alpha=0.5$ at the bottom half section ($y<0$), and from $\alpha=0.5$ to $\alpha=1$ at the top half section ($y>0$). However, the $\bm{\mathrm{'linear'}}$ curve only ranges from $\alpha=0.3$ to $\alpha=0.6$ (top half section) and from $\alpha=0.6$ to $\alpha=0.9$ (bottom half section) (see SM~\S3 for a more detailed definition of these distributions). This implies that the $\bm{\mathrm{'sine'}}$ curve possesses stronger asymmetry than $\bm{\mathrm{'linear'}}$ curve. Combining this implication with the observation that $h_c$ obtained for the $\bm{\mathrm{'sine'}}$ distribution is always larger than the $\bm{\mathrm{'linear'}}$ distribution (see Table~(\ref{tab: 2})), we verify the positive correlation between $h_c$ and the order distribution asymmetry.

        \begin{figure}[ht!]
            \centering
            \begin{subfigure}[b]{0.49\linewidth}
                \includegraphics[width=\linewidth]{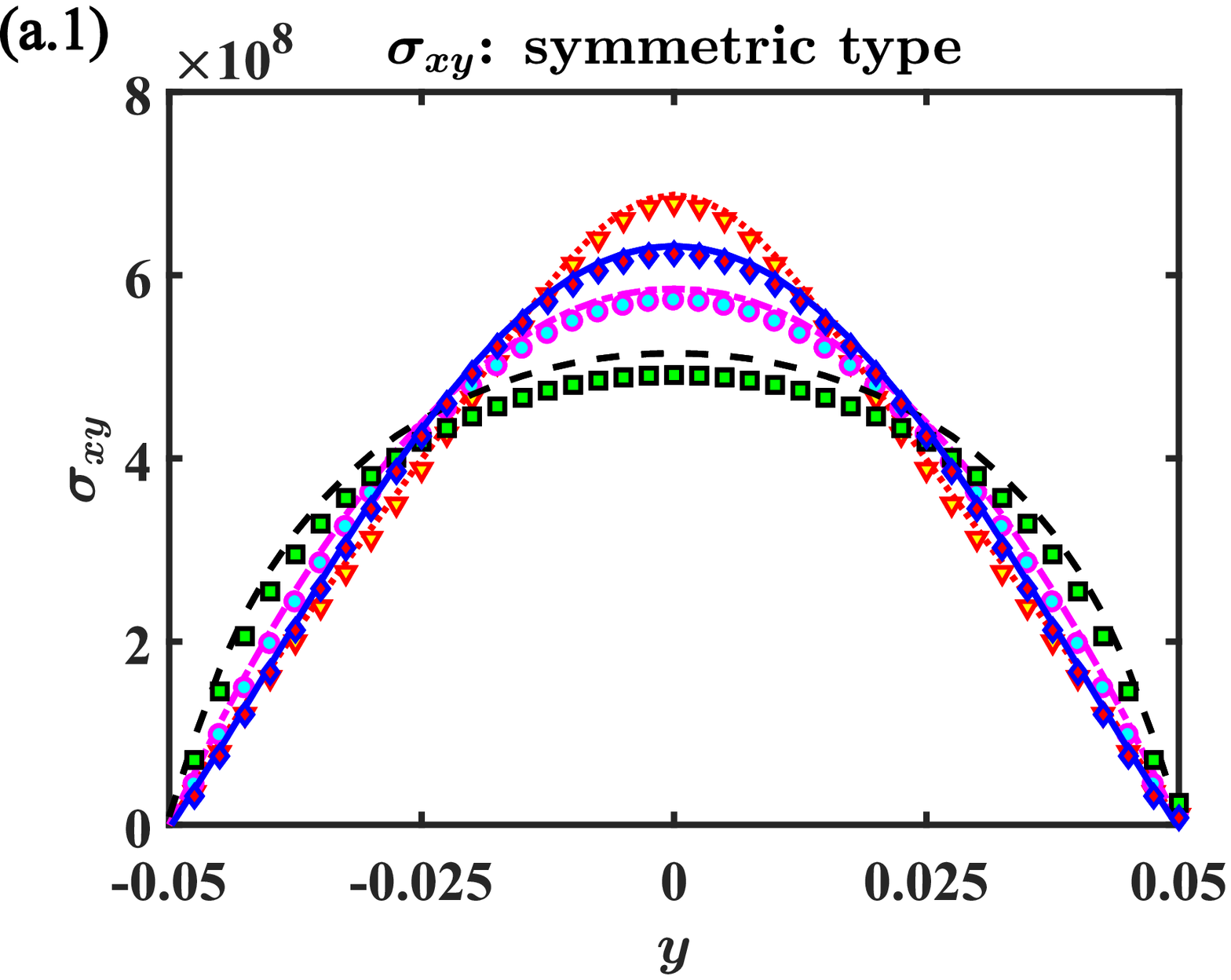}
            \end{subfigure}
            \begin{subfigure}[b]{0.49\linewidth}
                \includegraphics[width=\linewidth]{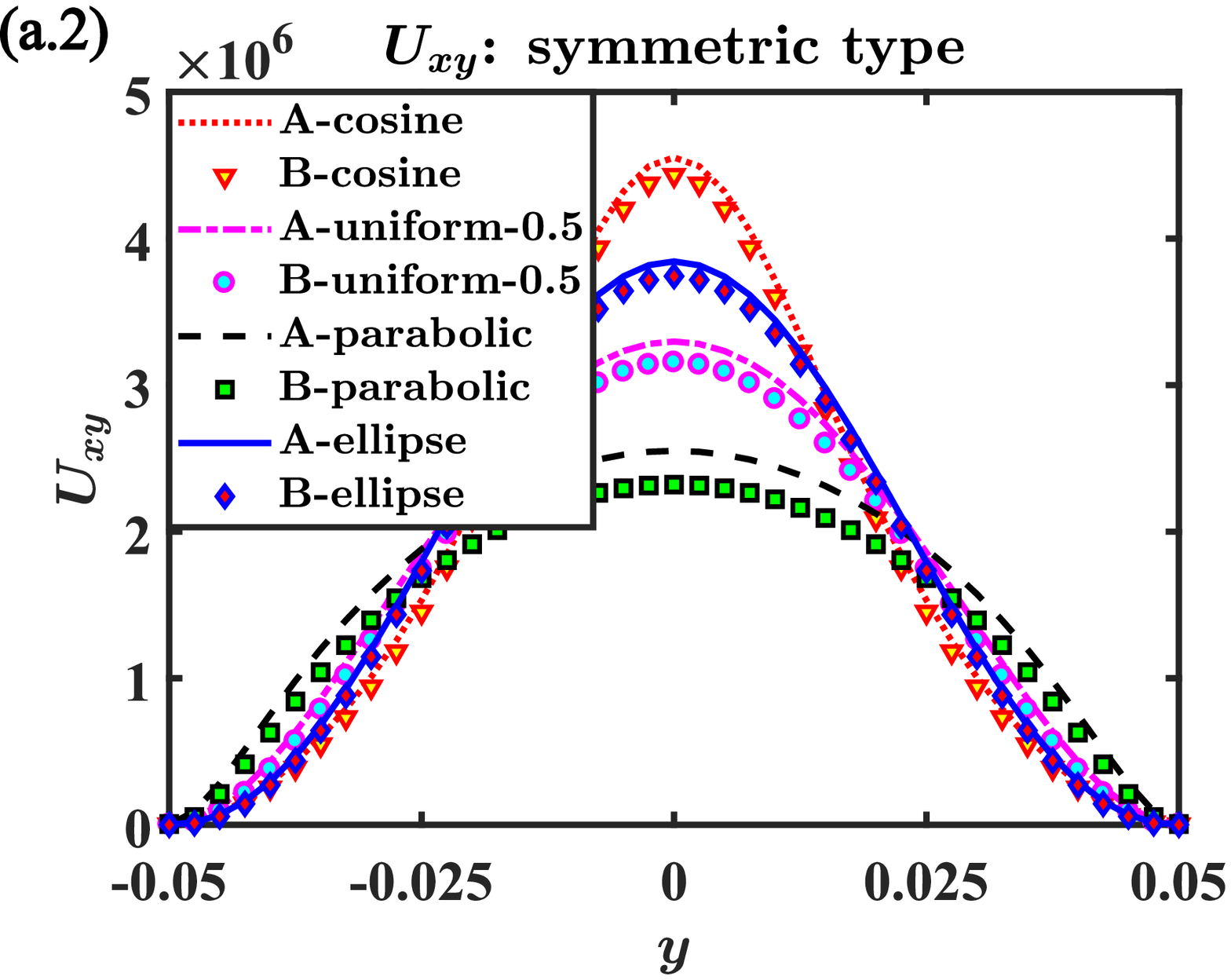}
            \end{subfigure}
            \begin{subfigure}[b]{0.49\linewidth}
                \includegraphics[width=\linewidth]{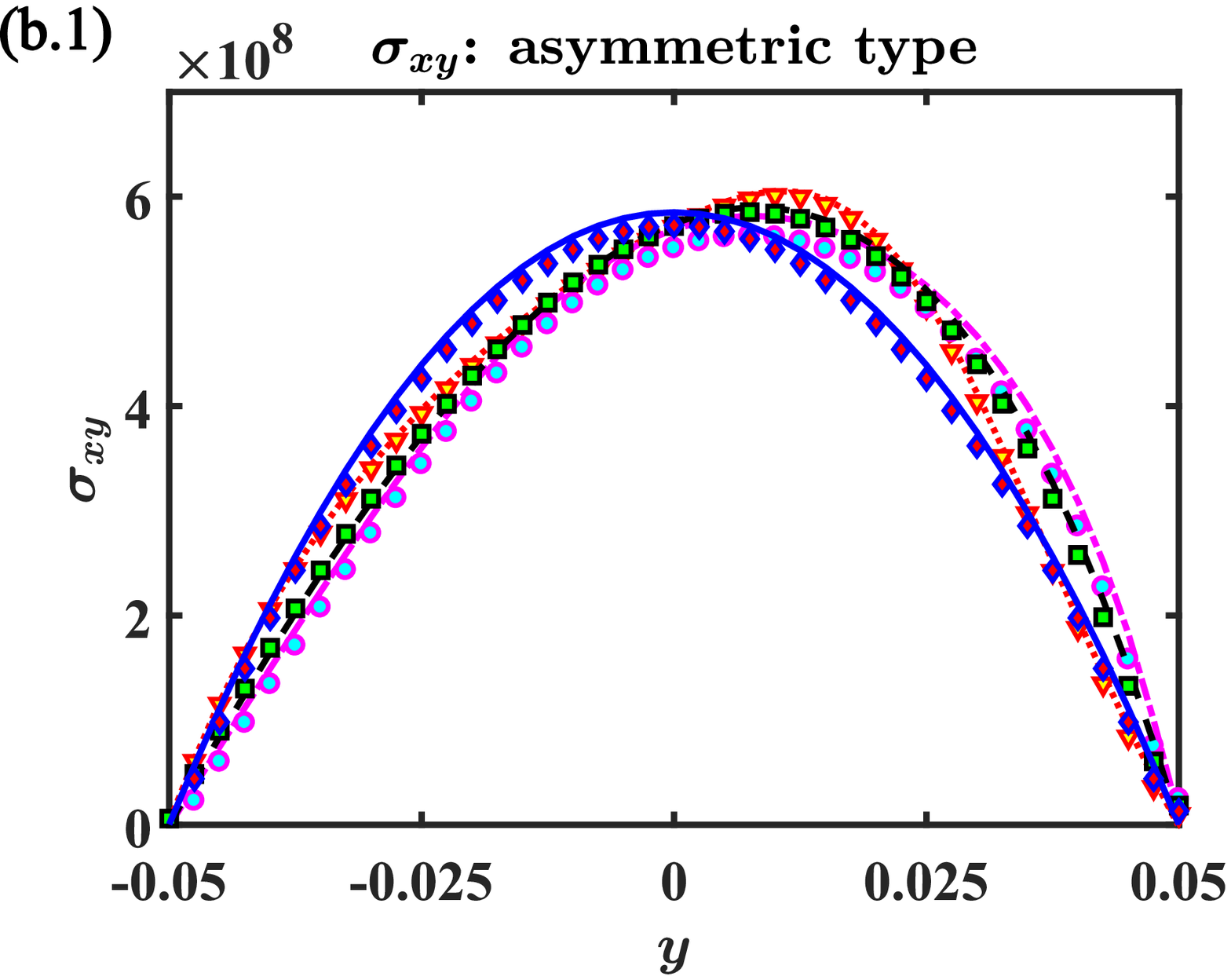}
            \end{subfigure}
            \begin{subfigure}[b]{0.49\linewidth}
                \includegraphics[width=\linewidth]{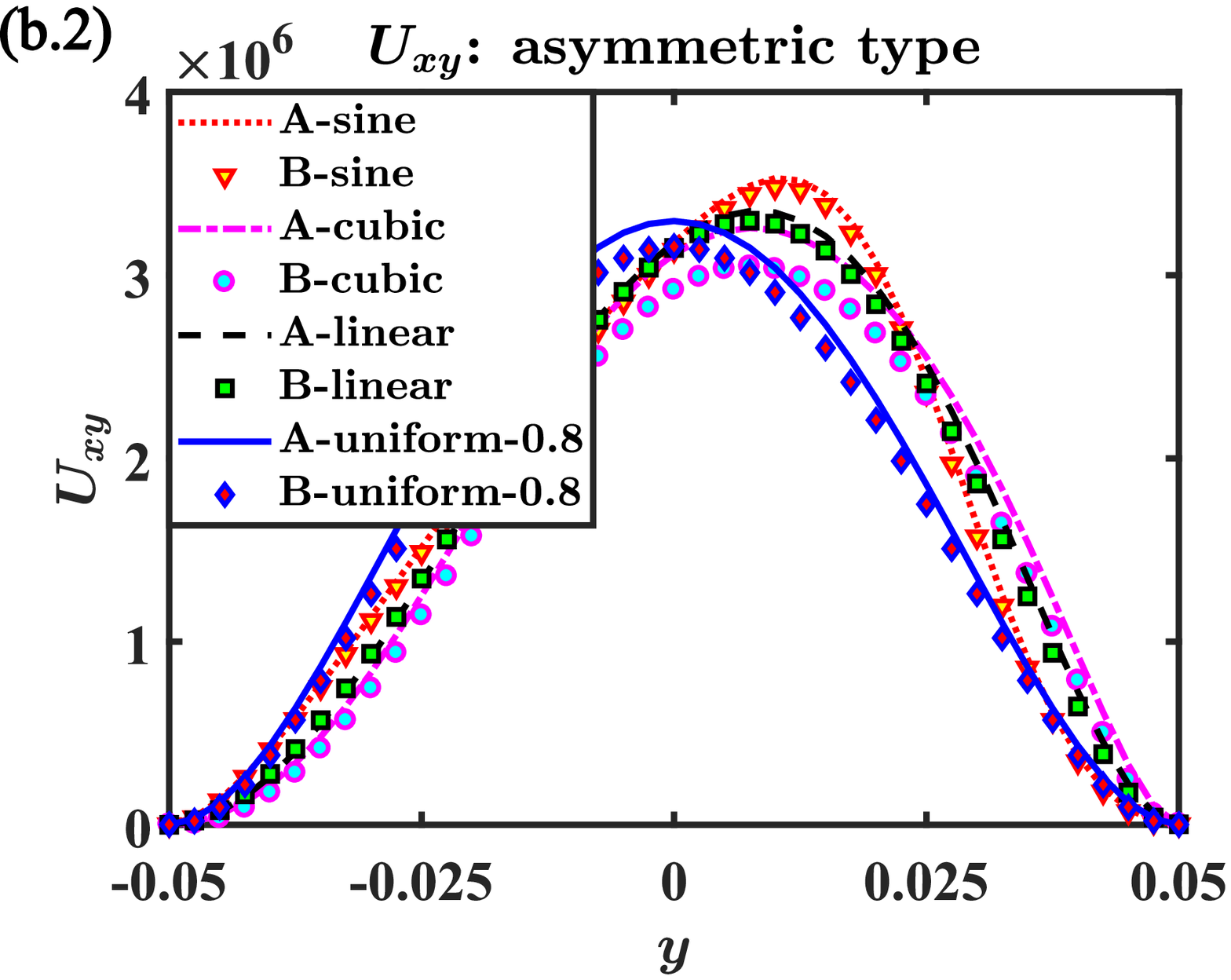}
            \end{subfigure}
            \caption{Distribution of shear stress $\sigma_{xy}$ and shear energy $U_{xy}$ along the cross section of the nonlocal beam at $x=0.3$m: (a.1) and (a.2) show $\sigma_{xy}$ and $U_{xy}$ with symmetric types distribution of order, respectively; (b.1) and (b.2) show $\sigma_{xy}$ and $U_{xy}$ with asymmetric types distribution of order, respectively. Note that $\bm{\mathrm{Problem~2}}$ is used for the simulation.}
            \label{fig: sigma_U_xy}
        \end{figure}

        \item While a strong correlation between $h_c$ and the order distribution $\alpha^{\textrm{b}}(y)$ can be explicitly interpreted from the physical perspective, the relation between $\chi$ and $\alpha^{\textrm{b}}(y)$ is relatively less prominent. Recall that in local Timoshenko beam theory, $\chi=5/6$ is typically chosen to account for the parabolic distribution of shear stress over the entire homogeneous rectangular-shaped cross section~\cite{timoshenko1955strength}. When considering the heterogeneous distribution of $\alpha^{\textrm{b}}(y)$, nonlocal properties over the cross section are not homogeneous anymore and as a result, affect the classical parabolic distribution of shear stress, and hence $\chi$. To study how $\chi$ actually changes with $\alpha^{\textrm{b}}(y)$, we analyze the transverse distribution of shear stress in nonlocal beams. 
        
        Fig.~\ref{fig: sigma_U_xy} shows the detailed distribution of shear stress $\sigma_{xy}$ and shear energy $U_{xy}$ in different simulations. We observe that both $\sigma_{xy}$ and $U_{xy}$ are closely related to the symmetry properties of $\alpha^{\textrm{b}}(y)$. For different types of symmetry of the $\alpha^{\textrm{b}}(y)$ distribution, $\sigma_{xy}$ and $U_{xy}$ are also symmetrically distributed but vary in gradient and peak value; for different asymmetric types of $\alpha^{\textrm{b}}(y)$, curves of $\sigma_{xy}$ and $U_{xy}$ are distorted and not symmetric anymore. Combining the simulation results in Fig.~\ref{fig: sigma_U_xy} together with Fig.~\ref{fig: alpha}, it appears that the material strength plays an important role in understanding the relation between $\alpha^{\textrm{b}}(y)$ and the shear stress (or shear energy). Recall that when $\alpha^{\textrm{b}}(y)$ increases, the material at the transverse position $y$ becomes stiffer and thus generates greater shear stress (or energy). Specifically, for the $\bm{\mathrm{'cosine'}}$ order distribution, since $\alpha^{\textrm{b}}(y)$ has the largest value at $y=0$, the material at $y=0$ is stiffer than the material at any other transverse position. In this regard, under the $\bm{\mathrm{'cosine'}}$ distribution the material at $y=0$ should store a greater amount of shear energy compared to a material with other types of $\alpha^{\textrm{b}}(y)$ distributions (that are softer at $y=0$); examples include materials with $\bm{\mathrm{'uniform\text{-}0.5'}}$ and $\bm{\mathrm{'parabolic'}}$ types of distributions. This analysis is consistent with the simulation results presented in Fig.~\ref{fig: sigma_U_xy}(a) where the shear stress and shear energy curves with $\alpha^{\textrm{b}}(y)$ in $\bm{\mathrm{'cosine'}}$ distribution at $y=0$ are always steeper and show higher peak values than their counterparts with $\alpha^{\textrm{b}}(y)$ in $\bm{\mathrm{'uniform\text{-}0.5'}}$ and $\bm{\mathrm{'parabolic'}}$ distributions.
        
        Given that both theoretical analysis and simulation results have confirmed the close connection between shear stress and shear energy distributions with $\alpha^{\textrm{b}}(y)$, the shear correction coefficient $\chi$ should also vary with different order distributions. As stated in~\cite{timoshenko1955strength}, $\chi$ is introduced to average the inhomogeneous shear stress distribution in beam problems. In other words, $\chi$ can be considered as an effective measure that represents the degree of uniformity with respect to shear stress distributions (similar to the variance defined in probability theory). According to the explicit formulations in Eqs.~(\ref{eq: sigma-Timoshenko},\ref{eq: chi-2}), as $\chi$ decreases the relative inconsistency between the uniform shear stress $\Tilde{\sigma}_{xy}$ and the real inhomogeneous shear stress ${\sigma}_{xy}$ increases, indicating that the distribution of ${\sigma}_{xy}$ is more dispersed. Simulation results in Fig.~\ref{fig: sigma_U_xy} and Table~(\ref{tab: 2}) further support this observation. Specifically, we note that for all simulation results with symmetric types of $\alpha^{\textrm{b}}(y)$ in Fig.~\ref{fig: sigma_U_xy}(a), $\sigma_{xy}$ and $U_{xy}$ curves with $\alpha^{\textrm{b}}(y)$ in $\bm{\mathrm{'cosine'}}$ are the steepest and highest, or alternatively, the most inhomogeneous. Correspondingly, curves with $\alpha^{\textrm{b}}(y)$ in $\bm{\mathrm{'parabolic'}}$ are observed to be the least inhomogeneous. As an effective measure of shear stress and shear energy variations, $\chi$ associated with $\bm{\mathrm{'cosine'}}$ distributions should be the smallest while $\chi$ associated with $\bm{\mathrm{'parabolic'}}$ distributions should be the largest. The variation of $\chi$, is found to be consistent with simulation results provided in Table~(\ref{tab: 2}). With the relationship between $\chi$ and $\alpha^{\textrm{b}}(y)$ being fully characterized, we remark that $\chi$ can be regarded as a key parameter to better capture the internal variation of shear effects and herein enhances the DO-NTBM ability to model heterogeneous nonlocal properties at the micro scales.
    \end{itemize}
\end{enumerate}

\subsection{Convergence analysis and computational cost}\label{ssec: convergence analysis}
In this section, we analyze the computational performance of nonlocal beam formulations. Given that the two approaches differ in modeling dimensions (2D for DO-ANET and 1D DO-NTBM) and numerical algorithms (see SM~\S2), the numerical performance (e.g. convergence and computational cost) are expected to be different. Moreover, computational efficiency is also strongly affected by nonlocal effects. The analysis will focus on the following two aspects: 1) the difference of computational performance between the two modeling approaches, and 2) the relation between computational efficiency and nonlocal effects. Fig.~\ref{fig: stiffness_matrix} and Fig.~\ref{fig: time_convergence} present detailed results in terms of the stiffness matrix structure, computational time cost, and convergence. A detailed analysis of these results leads to the following observations and conclusions:

\begin{figure}[ht!]
    \centering
    \includegraphics[width=0.95\textwidth]{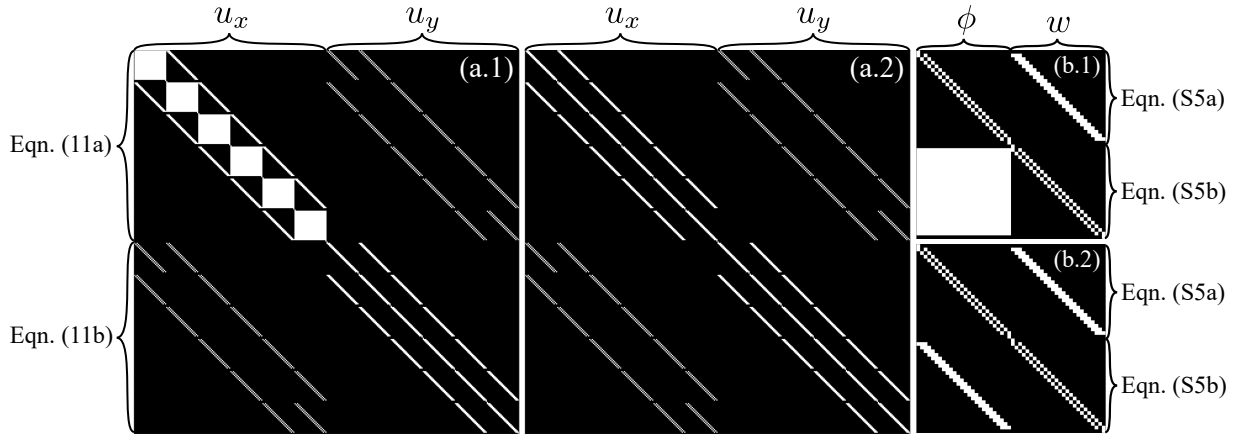}
    \caption{Structure of the stiffness matrices in which zero and nonzero entries are indicated by black color and white color, respectively. (a.1) and (a.2) show the stiffness matrices obtained by the DO-ANET for either the local or the nonlocal beams, respectively. Columns that correspond to $u_x$ and $u_y$ (also rows that correspond to the governing equations in Eq.~(\ref{eq: governing_eqn_beam}a) and Eq.~(\ref{eq: governing_eqn_beam}b)) are labeled explicitly in (a.1) and (a.2). (b.1) and (b.2) show the stiffness matrices obtained by the DO-NTBM for either the local or the nonlocal beam, respectively. Similarly, columns that correspond to $\phi$ and $w$ (also rows that correspond to Eq.~(S5a) and Eq.~(S5b)) are labeled explicitly in (b.1) and (b.2). It can be observed that the blocks of the stiffness matrix are not sparse for equations (such as Eq.~(\ref{eq: governing_eqn_beam}a) and Eq.~(S5b)) involving nonlocal formulations. These matrices were generated assuming a beam geometry $L \times h = 1\mathrm{m} \times 0.2\mathrm{m}$ and discretization $N_x \times N_y = 26 \times 6$.}
    \label{fig: stiffness_matrix}
\end{figure}

\begin{enumerate}[leftmargin=*]
    \item[$\bm{(1)}$] The computational time grows proportionally to the third power of the number of elements (characterized by $N_x$ or $N_y$) for DO-ANET and quadratic growth for DO-NTBM. The computational time cost difference between the two nonlocal approaches stems from the different modeling dimensions. While the DO-ANET requires a full 2D discretization with total $N_x \times N_y$ mesh points, the DO-NTBM accounts for the heterogeneities in the $y$-direction into a single DO operator and only requires 1D discretization with $N_x$ mesh points. Specifically, consider a given nonlocal beam with total mesh points $N_x \times N_y~(N_{\alpha})$ and the aspect ratio $r = L/h = N_x / N_y$, the total degrees of freedom (DOF) required in each approach can be evaluated as:
    \begin{equation}\label{eq: DOF}
    \begin{aligned}
        \textrm{DOF}_{\textrm{ANET}} &= \underbrace{N_x}_{P_1} \times \underbrace{(N_x \times N_y)}_{Q_1} = r^2N_y^3 = \mathcal{O}(N_y^3) \\
        \textrm{DOF}_{\textrm{NTBM}} &= \underbrace{N_x}_{P_2} \times \underbrace{N_x}_{Q_2} = r^2N_y^2 = \mathcal{O}(N_y^2) \\
    \end{aligned}
    \end{equation}
    where $Q_1$ and $Q_2$ denote the total number of mesh points in DO-ANET and DO-NTBM, respectively. In both models, each mesh point  is interconnected with other $N_x-1$ mesh points due to the nonlocal interaction spanning the entire axial direction. $P_1$ and $P_2$ denote the total number of mesh points that interact with each other within the nonlocal horizon (recall that in Fig.~\ref{fig: Overview} we define the nonlocal horizon as the length of the whole beam). We remark that, the total DOF for DO-ANET (see $\textrm{DOF}_{\textrm{ANET}}$ in Eq.~(\ref{eq: DOF})) and DO-NTBM (see $\textrm{DOF}_{\textrm{NTBM}}$ in Eq.~(\ref{eq: DOF})) are on the order of the third and second power of $N_y$, respectively.
    
    The above analysis of the total number of DOF in each model can be further substantiated based on simulation results. Fig.~\ref{fig: stiffness_matrix} shows the structure of the stiffness matrices. We observe that due to the nonlocal interactions, the stiffness matrices obtained by both the DO-ANET and the DO-NTBM contain nonzero blocks and are not sparse. The presence of these nonzero blocks, compared with other sparse distribution of nonzero entries, is the prominent factor that leads to high computational costs in nonlocal problems. Note that, using the discretization $N_x \times N_y= 26 \times 6$, the stiffness matrix in DO-ANET has $N_y=6$ nonzero blocks in Fig.~\ref{fig: stiffness_matrix}(a.2), while the stiffness matrix in DO-NTBM has only one nonzero block in Fig.~\ref{fig: stiffness_matrix}(b.2). The difference between the number of nonzero blocks is in agreement with the theoretical formulation in Eq.~(\ref{eq: DOF}). The same characteristic can be further justified by the evaluation of the computational time in Fig.~\ref{fig: time_convergence}(a). We observe that the DO-ANET curve (using direct solver, see the squared black line) shows $\mathcal{O}(N_y^3)$ growth pattern and the DO-NTBM curve (see the dotted red line) shows $\mathcal{O}(N_y^2)$ growth pattern. This observation is not strictly valid for larger scale problems, such as when $N_y > 60$, because both approaches require the use of an iteration process (see detailed algorithms in SM~\S2) and lead either to higher growth order (for DO-ANET) or to slight fluctuation of computational time cost (for DO-NTBM). In general, we conclude that both theoretical and numerical results indicate the superior computational efficiency of DO-NTBM when modeling multiscale nonlocal beams. 
    
    The necessity of developing more efficient nonlocal modeling approaches can be further justified by analyzing the effects of nonlocality on the computational efficiency. As it shows in Fig.~\ref{fig: time_convergence}(a), the computational cost of numerical simulations obtained either by nonlocal or local DO-ANET follow cubic and quadratic growth, respectively. It emerges that the different computational cost between the two cases is the direct result of nonlocal effects. For the nonlocal case, each mesh point is affected by all the other $N_x-1$ points located in the axial direction due to nonlocal interactions; for the local case, each mesh point only interacts locally with its neighbor points. The increment of the total number of interacting points due to nonlocality can be observed in Fig.~\ref{fig: stiffness_matrix}. In comparison with the sparse nature of the local stiffness matrices (see subfigures (a.2) and (b.2)), the nonlocal stiffness matrices (see subfigures (a.1) and (b.1)) contain nonzero blocks and are non-sparse. As a consequence, this lack of sparsity of the nonlocal stiffness matrices, levels up computational complexity and increases computational costs. In this regard, it is important to develop efficient modeling approaches, such as DO-NTBM, in order to mitigate the effect of nonlocality on computational efficiency and facilitate the analysis of nonlocal structural mechanics.

    \begin{figure}[ht]
        \centering
        \begin{subfigure}[b]{0.49\linewidth}
            \includegraphics[width=\linewidth]{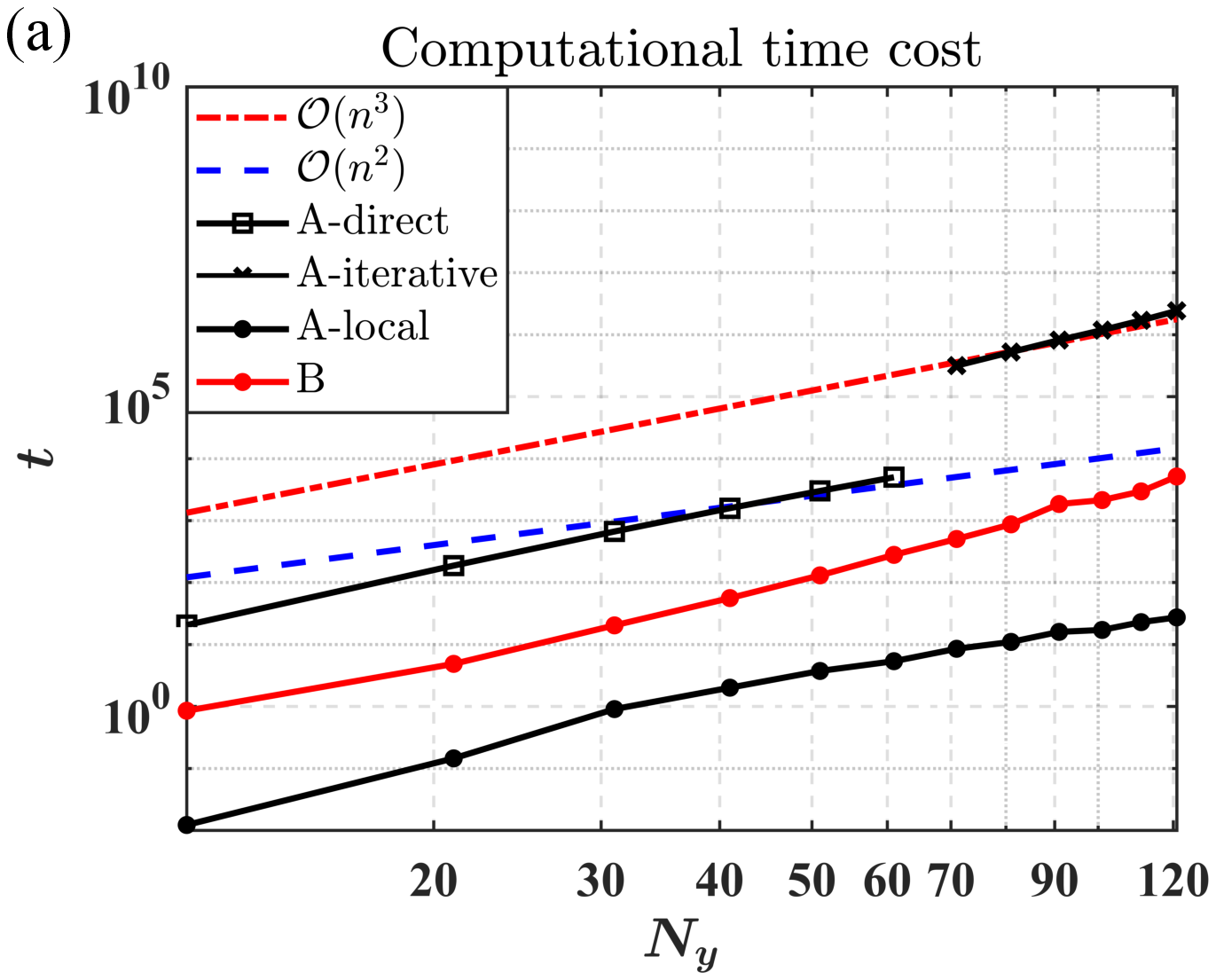}
        \end{subfigure}
        \begin{subfigure}[b]{0.49\linewidth}
            \includegraphics[width=\linewidth]{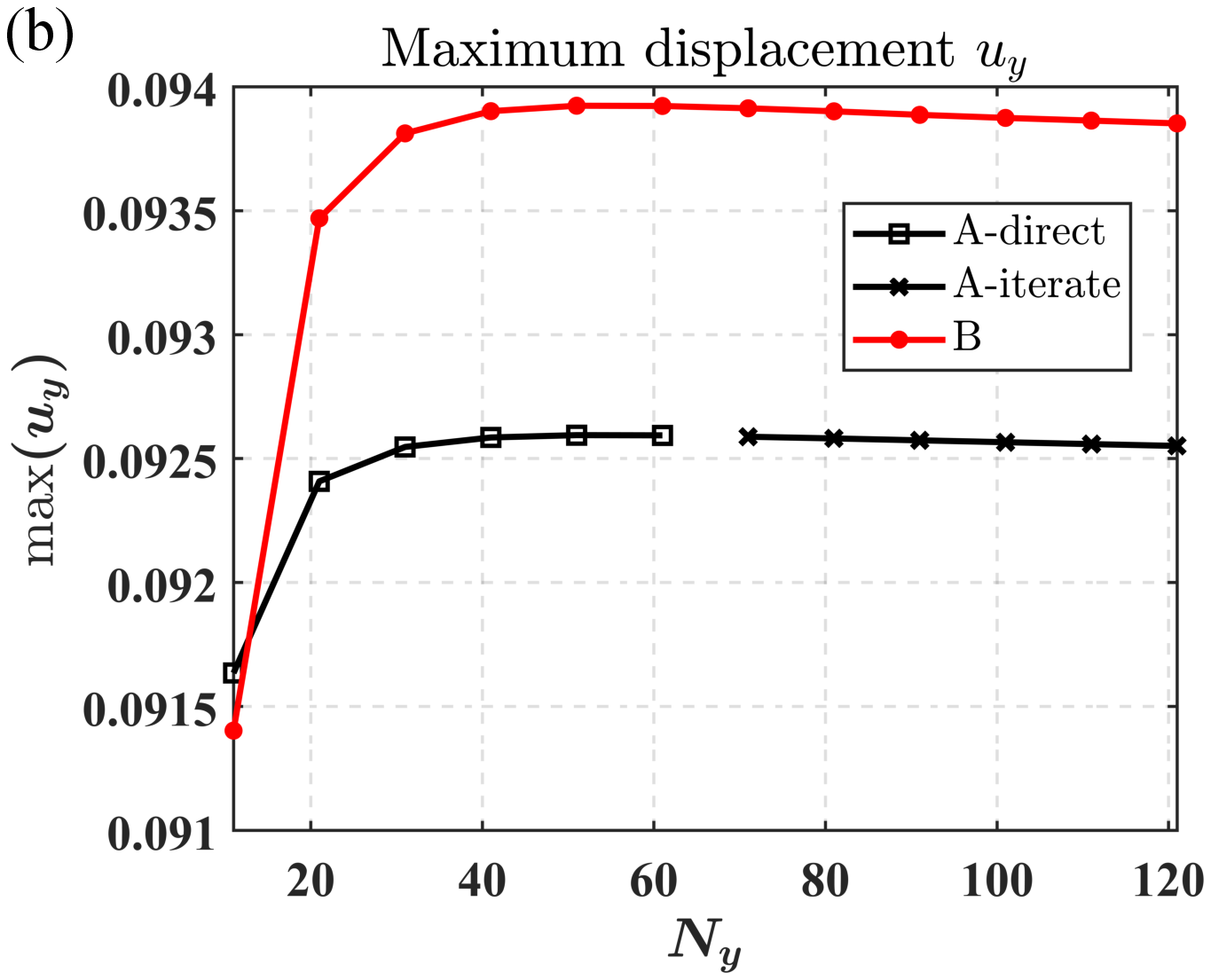}
        \end{subfigure}
        \caption{Simulation results showing (a) computational time and (b) convergence performance. Specifically, simulation results presented in (a) consist of: 1) computational time for the nonlocal beams solved using the direct DO-ANET solver (black squared line), the iterative DO-ANET solver (black cross line), and the DO-NTBM solver (red dotted line), and 2) computational time for the local beams solved using the direct DO-ANET solver (black dotted line). Convergence plot in (b) shows the maximum value of $u_y$ for simulations performed using the direct DO-ANET solver (black squared line), iterative DO-ANET solver (black cross line), and DO-NTBM solver (red dotted line). $\bm{\mathrm{'A'}}$ and $\bm{\mathrm{'B'}}$ in both legends stand for DO-ANET and DO-NTBM, respectively. $\bm{\mathrm{Problem~2}}$ and $\bm{\mathrm{'linear'}}$ distribution of $\alpha^{\mathrm{b}}(y)$ are used for the nonlocal beam simulations. Direct and iterative nonlocal DO-ANET simulations were performed using MATLAB parallel computing toolbox with 50 processor cores and 120 processor cores, respectively. Local DO-ANET and DO-NTBM simulations were tested using serial computations based on a 1 processor core. All the numerical tests were performed on a high power computing cluster using AMD Rome CPU @ 2.0GHz.}
        \label{fig: time_convergence}
    \end{figure}

    \item[$\bm{(3)}$] Both the DO-ANET and the DO-NTBM show good convergence. Fig.~\ref{fig: time_convergence}(b) presents detailed simulation results of convergence, in terms of the maximum transverse displacement $u_y$ versus the number of mesh points in the $y$-direction ($N_y$). We observe that for $N_y < 60$, $u_y$ computed by both direct DO-ANET and DO-NTBM converge fast to certain values (roughly 0.0926 for the former case and 0.0939 for the latter case). For $N_y > 60$, we observe a slight decay of maximum $u_y$ in both iterative DO-ANET and DO-NTBM. A plausible explanation for this unconventional behavior, along with the higher growth order and the fluctuation of time cost in Fig.~\ref{fig: time_convergence}, is that both simulations require the use of iterative methods. In addition, for the iterative DO-ANET the total number of degrees of freedom in the final stiffness matrix increases drastically when $N_y$ increases (in view of its 2D nature). Taking $N_x \times N_y = 1200 \times 120$ as an example, the final stiffness matrix is non-sparse (see Fig.~\ref{fig: stiffness_matrix}(a.2)) and has more than $N_x \times N_x \times N_y \approx 1.75 \times 10^{8} $ nonzero entries. Such a large non-sparse stiffness matrix could possibly lead to computational complexity and even numerical error. 
\end{enumerate}

Based on the above discussions focusing on the elastostatic analyses as well as on the computational aspects of nonlocal beams, we conclude that the simulation results obtained by both the DO-ANET and DO-NTBM are in good agreement. Both theoretical and numerical analyses presented above point towards the superior capability of DO-NTBM to model complex multiscale nonlocal beams. More specifically, DO-NTBM can accurately predict the response of nonlocal beams while significantly reducing the computational cost. We also conclude that the multiscale nonlocal information and its effect on the beam elastic properties are well retained via the DO-NTBM formulation. At the macroscopic scales, the distribution of transverse displacement predicted by DO-NTBM is consistent with the results obtained by the 2D DO-ANET. At the microscopic scale, the overall transverse distribution of stress and shear energy reproduced by the DO-NTBM is also well consistent with the 2D solutions obtained by the DO-ANET. In general, the DO-NTBM provides an effective and computationally efficient approach to model the system at the macro scales while still retaining the information from the micro scales.
    
\section{Conclusions}\label{sec: conclusion}
This study presented a one-dimensional (1D) distributed-order (DO) nonlocal Timoshenko beam (DO-NTBM) formulation derived from a theoretical framework of two-dimensional (2D) anisotropic nonlocal elasticity theory (DO-ANET). The approach was applied to model 2D elastic beams with uniaxial nonlocality that is heterogeneously distributed over the transverse (thickness) direction. Distributed-order (DO) operators with a spatially-dependent fourth-order strength-function tensor were introduced in the nonlocal elasticity formulation to account for both nonlocal anisotropy and heterogeneity. The DO-NTBM accounts for the heterogeneous nonlocal information across the thickness direction by leveraging the unique multiscale properties of DO operators. The resulting DO Timoshenko beam formulation allowed accounting for nonlocal effects while simultaneously reducing the problem dimensions and significantly scaling down the computational costs. Numerical simulations were performed to validate the effectiveness of the proposed modeling approaches. Both methodologies were applied to simulate multiscale nonlocal beams under various conditions. It was observed that the overall transverse beam displacements predicted by the two approaches were generally in good agreement. Both approaches captured the well-known material softening due to the nonlocal effects, ultimately leading to larger transverse displacements. The detailed transverse distribution of mechanical field quantities predicted by the two models, such as stress and shear energy, was also shown to be consistent. The consistency between simulation results at both macroscopic and microscopic scales, revealed the unique multiscale nonlocal characteristics of the proposed DO-NTBM. Investigations on the computational costs further justified the superior computational efficiency of DO-NTBM. In conclusion, the results presented in this study highlighted several unique features of DO operators for applications to multiscale nonlocal problems and suggested that this approach could provide a solid foundation to develop accurate and efficient computational platforms to simulate complex multiscale nonlocal systems.\newline

\noindent\textbf{Data Availability.} All the necessary data and information required to reproduce the results are available in the paper and the supplementary information document.\newline

\noindent\textbf{Acknowledgements.} The authors gratefully acknowledge the financial support of the National Science Foundation under grants MOMS \#1761423, DCSD \#1825837, and the Defense Advanced Research Project Agency under grant \#D19AP00052. Any opinions, findings, and conclusions or recommendations expressed in this material are those of the author(s) and do not necessarily reflect the views of the National Science Foundation. The content and information presented in this manuscript do not necessarily reflect the position or the policy of the government. The material is approved for public release; distribution is unlimited.\newline


\noindent\textbf{Competing Interests.} The authors declare that there are no competing interests.

\bibliographystyle{naturemag}
\bibliography{Report}
\end{document}